\documentclass[11pt]{article}
\usepackage{amssymb}
\usepackage{graphicx}
\usepackage{layout}
\usepackage{bm}

\begin{document}

\author{Guido Rizzi,$^\S$ Matteo Luca Ruggiero,$^\S$ Alessio Serafini$^\ast$
\\ \\
\small
$^\S$ Dipartimento di Fisica, Politecnico di Torino and INFN, Sezione di Torino\\
\small $^\ast$ Dipartimento di Fisica ``E.R. Caianiello'', Universit\`{a} di Salerno\\ \\
\small E-mail: guido.rizzi@polito.it, matteo.ruggiero@polito.it,
serale@sa.infn.it}
\title{Synchronization Gauges and the Principles of Special Relativity}
\maketitle

\begin{abstract}
The axiomatic bases of Special Relativity Theory (SRT) are
thoroughly re-examined from an operational point of view, with
particular emphasis on the status of Einstein synchronization in the light
of the possibility of arbitrary synchronization procedures in inertial
reference frames. Once correctly and explicitly phrased, the principles of
SRT allow for a wide range of `theories' that differ from the standard SRT only
for the difference in the chosen synchronization procedures, but are
wholly equivalent to SRT in predicting empirical facts.
This results in the introduction, in the full background of SRT,
of a suitable synchronization gauge. A complete hierarchy of synchronization
gauges is introduced and elucidated, ranging from the useful Selleri
synchronization gauge (which should lead, according to Selleri, to a
multiplicity of theories alternative to SRT) to the more general
Mansouri-Sexl synchronization gauge and, finally, to the even more general
Anderson-Vetharaniam-Stedman's synchronization gauge. It is showed that all
these gauges do not challenge the SRT, as claimed by Selleri, but simply
lead to a number of formalisms which leave the geometrical structure of
Minkowski spacetime unchanged. Several aspects of fundamental and applied
interest related to the conventional aspect of the synchronization choice
are discussed, encompassing the issue of the one-way velocity of light on
inertial and rotating reference frames, the GPS's working, and the recasting
of Maxwell equations in generic synchronizations. Finally, it is showed how
the gauge freedom introduced in SRT can be exploited in order to give a
clear explanation of the Sagnac effect for counter-propagating matter beams.
\end{abstract}
\normalsize


\section{Introduction}

\subsection{The issue of conventionality of synchronization and one-way
speed of light: a bit of history}

As well known, in Special Relativity Theory (SRT) it is assumed that
``clocks can be adjusted in such a way that the propagation velocity of
every light ray in vacuum - measured by means of these clocks - becomes
everywhere equal to a universal constant c, provided that the coordinate
system is not accelerated'' \cite{Einstein1} (Einstein 1907). Clocks
adjusted in such a way define the so-called ``Einstein synchronization''.
Any inertial reference frame (IRF) turns out to be optically isotropic if,
and only if, Einstein synchronization is adopted ``\textit{by stipulation''.%
\footnote{%
In his popular exposition of the theory (first ed. 1916), Einstein
explicitly recognized that the isotropic propagation of light is ``neither a
\textit{supposition} nor a \textit{hypotesis} about the physical nature of
light, but a \textit{stipulation'' \cite{Einstein2}}} }Anyway, Einstein
points out that such a stipulation, although arbitrary on a purely logical
viewpoint (as a matter of fact, we have empirical access only to the
round-trip average speed of light), is not arbitrary on the physical
viewpoint since it provides a symmetric and transitive relationship. As a
consequence, the standard formulation of SRT gives for granted that every
light ray actually propagates isotropically; consequently, simultaneity is
indeed frame-dependent.

However, some authors rejected the thesis that Einstein
synchronization (with all its implications, like the isotropic
propagation of light) is the only choice enforced by experimental
data\footnote{Of course this is not literally Einstein's thesis
(see the previous footnote), but a thesis firmly incorporated in
the standard - traditional - formulation of SRT}, and seriously
considered the possibility of postulating an anisotropic
propagation of light in the theoretical context of SRT. Of course,
a possible anisotropic propagation of light could be accounted for
only on the basis of a possible synchronization gauge consistent
with all the empirical observations, provided that any synchrony
choice belonging to
this gauge provides a symmetric and transitive relationship among events.%
\footnote{%
A synchronization gauge is a (very particular) group of
transformations, internal to the theory, which leave the
observables unchanged (''saving the phenomena'')} This viewpoint,
leading to the thesis of ``conventionality of simultaneity'', has
been discussed extensively, often from a philosophical standpoint,
by many authors, in particular by Reichenbach \cite{reichenbach2}
and Gr\"{u}nbaum \cite{grunbaum}. Although starting from different
points of view, these authors agree on the idea that the only
nonconventional basis for claiming that two distinct events are
not simultaneous would be the possibility of a causal influence
connecting the events; as a consequence any self-consistent
definition of simultaneity between ``spatially separated'' events
is, in principle, allowed in Minkowski spacetime.

Nowadays the thesis of ``conventionality of simultaneity'' is a
talking-point among philosophers of science, but seems not
particularly exciting among relativistic people, who are generally
satisfied with the
standard formulation of SRT. As a matter of fact, most relativistic authors%
\footnote{Let us limit ourselves to quote some of the most recent
claims by Bergia \cite{bergia}, Tartaglia\cite{tartaglia_libro},
Sorge\cite{sorge_libro}.} sustain the standard traditional
viewpoint, according to which it can be proved, \textit{on
experimental grounds}, that the one-way velocity of light, in any
inertial reference frame (IRF) and in any direction, is an
universal constant. Let us quote some examples of standard (i.e.
anti-conventionalist) approaches.\bigskip

In the late 1960's Ellis and Bowman \cite{Ellis-Bowman}, after careful
synchronization of clocks by slow transport, argue that, although consistent
nonstandard synchronization does not appear totally ruled out, there are
sound physical reasons for preferring standard Einstein synchronization.
Their conclusion is: ''the thesis of the conventionality of distant
simultaneity espoused particularly by Reichenbach and Gr\"{u}nbaum is thus
either trivialized or refuted''.

Malament\cite{malament} argues that standard synchrony is the only
simultaneity relation that can be defined, relative to a given
IRF, from the relation of
(symmetric) causal connectibility. Let this relation be represented by $%
\kappa $, and let the statement that events p and q are simultaneous be
represented by S(p,q). Under suitable formal conditions (in particular S
should be an equivalence relation definable from $\kappa $ in any IRF),
Malament's theorem asserts that there is one and only one S, namely the
relation of standard synchrony. More explicitly, Malament's theorem shows
that the standard simultaneity relation is the only nontrivial simultaneity
relation definable in terms of the causal structure of the Minkowski
spacetime of SRT.

Following Malament, Friedman \cite{friedman} (1983) claims that any
non-Einstein synchrony choice entails a denial of the Minkowskian structure
of spacetime (which amounts to a denial of SRT).

More recently, Bergia and Valleriani \cite{bergia}, \cite{bergia-valleriani}
claim that the Einstein synchrony choice is not conventional at all, but is
forced by some experimental evidences. In particular, they quote two
astonishing evidences: the one coming from the faultless performance of the
world wide network of atomic Einstein-synchronized clocks (i.e. clocks
synchronized by means of radio signals), and the one coming from faultless
performance of the Global Positioning System (GPS). The main argument is
that, in both cases, a possible anisotropic propagation of light should
cause some detectable delays, fortunately never observed, which could be
more than enough to obstruct the accurate performance of these devices. A
careful quantitative analysis, based on the estimated but never observed
delays, shows that the possible relative spread of the speed of light is
about $\Delta c/c\sim 1/300000\sim \allowbreak 3\cdot 10^{-6}$: more than
enough to rule out any alternative synchrony choice.

In the anti-conventionalist area, the issue of arbitrary
synchronizations has enjoyed much consideration in order to
suggest experimental tests of SRT. The approach is to use some
kind of ``test-theory'', i.e. a theoretical framework containing a
(suitably parameterized) family of theories in competition with
one other. The main characteristic of a test-theory is the
presence of a particular set of parameters whose numerical values
are specific of any specific theory to be tested in such a
framework. If the parameters are related to some observables, then
suitable experiments can single out the correct theory belonging
to the test-theory. Although none of the test-theory of SRT enjoy
the same status as the celebrated PPN test-theory of GRT, the
test-theory approach is promising, provided that the crucial
parameters are actually related to observable  (and not to
conventional) quantities: unfortunately, this is not always the
case.

The most popular test-theories are the ones of Mansouri-Sexl,
which yet contains a very important element of conventionality
(see later), and the ones reviewed by Clifford Will
\cite{will2001}, in a theoretical background which tries to test
the local Lorentz symmetry by measuring a possible difference
between the speed of electromagnetic radiation c and the limiting
speed of material test particles, chosen to be unity via a
suitable choice of units. According to such an approach, the
relevant parameter is $\delta \doteq c^{-2}-1$. Possible
deviations from the standard value zero would unveil a violation
of Lorentz symmetry, selecting a preferred universal rest frame:
presumably that of the cosmic background radiation, through which
we are moving with a velocity contained in the range 300-600 km/s.
Will quotes some selected tests of local Lorentz symmetry showing
the bounds on the parameter $\delta $. According to him, ``recent
advances in atomic spectroscopy and atomic timekeeping have made
it possible to test local Lorentz symmetry by checking the
isotropy of the speed of light using one-way propagation (as
opposed to round-trip propagation, as in the Michelson-Morley
experiment)''. Since the bounds on the parameter $\delta $ turn
out to be contained in the range $\left( 10^{-6},10^{-20}\right)
,$
depending on the experiment, the relative spread of the speed of light $%
\Delta c/c$ compatible with the experiments agrees with the one found by
Bergia \cite{bergia}. As a consequence, the one-way isotropic propagation of
light turns out to be the only possibility allowed by experiments.\footnote{%
In a previous paper \cite{will} Will goes into details, assuming (one
version of) the Mansouri-Sexl test-theory as a starting point. Some relevant
``one-way'' experiments, like the two-photon-absorption (TPA) experiment,
are described. It is stressed that the Mansouri-Sexl transformations from an
Einstein-synchronized IRF $\Sigma $ to an arbitrary-synchronized IRF $S$
embody one vector parameter $\mathbf{\varepsilon }$ which depends on the
synchrony choice in $S$, and 3 scalar parameters $a,b,d$ which do not depend
on this synchrony choice. That having said, Will shows that ``the outcome of
of physical experiments... is independent of synchronization''; yet ``the
TPA and other such one-way experiments do provide valid tests to possible
violations of SRT''. However odd, this is possible because ``those
violations are embodied in functional forms of $a,b,d$ that could differ
from those quoted above [the ones expected in SRT], not in the form of $%
\mathbf{\varepsilon }$ which is arbitrary an irrelevant''. The conclusion is
that ``the TPA and other such one-way experiments'' actually prove the
isotropy of the one-way velocity of light (with relative spread severely
bounded under $10^{-7}$) \textit{independently of any synchrony choice}.%
\textit{\ }We report here this result, but we acknowledge our
incomprehension since the one-way isotropic propagation of light should
imply the Einstein synchrony choice (and vice versa). We completely agree
with AVS \cite{avs} who point out that all parameters - $a,b,d$ included -
depend on the synchrony choice in $\Sigma $: this fact is somehow obscured
by the Einstein-synchrony choice in $\Sigma $, but becomes apparent if also $%
\Sigma $ is arbitrarily synchronized.}

These are some of the most outstanding results found by authors who believe
in the possibility of proving the isotropy of one-way propagation of light
by means of actual experiments. \bigskip

However, in spite of such strong and apparently conclusive claims,
some underground perplexities about the possibility of reliable
experimental tests of such a statement went through all along the
history of SRT, because of the inescapable entanglement between
remote synchronization and one-way velocity of light. After all, a
careful analysis (see f.i. \cite{avs},
\cite{selleri4,selleri3,selleri,selleri2}) seems to unveil that no
actual experiment, among the ones mentioned by the authors quoted
before, is a genuine ``one-way'' experiment. As a consequence, it
seems reasonable to suspect that the parameters appearing in the
test-theories mentioned before could actually be beyond the reach
of experimental knowledge: in other words, these parameters could
be ``conventional'' in the sense that their numerical values have
no effect on the output of any actual experiment.

Likewise, it seems reasonable to suspect that some claims
mentioned before could arise from circular arguments, in the sense
that the conclusions could be hidden, from the very beginning, in
the hypotheses. For instance, it can be shown \cite{RS_libro} that
synchronization of clocks by slow transport is fully equivalent to
Einstein synchronization, provided that the geometrical
structure of Minkowski spacetime is accepted once and for all.\footnote{%
Let us stress that accepting the geometrical structure of
Minkowski spacetime is equivalent to accepting SRT, regardless of
its formal look. Therefore we completely agree with the following
AVS \cite{avs} claim: ``any experimental divergence between
Einstein synchronization and slow clock transport would constitute
an experimental violation of special relativity''.} So that the
refutation of any non-Einstein synchrony choice, starting from
synchronization of clocks by slow transport, runs the risk of
being tautological, unless Minkowski spacetime is embedded in a
set of suitably parametrized (flat) spacetimes.

A major breakthrough occurred in 1977, when Mansouri and Sexl
\cite{mansouri} introduced a synchronization gauge consistent with
the experimental evidence of the constancy of the two-ways
velocity of light in any IRF. This is a historical cornerstone of
the conventionalist viewpoint. However, in the same year, Mansouri
and Sexl published three papers, and it is apparent that their
philosophical approach gradually evolved from the starting
position of acknowledging the conventionality of simultaneity to
the opposing position that each theory has associated with it an
empirically determinable synchrony choice. In particular, they
came to the conclusion that the one-way speed of light could be
empirically measured in the framework of their theory. In this way
the Mansouri-Sexl approach becomes a test-theory; as a matter of
fact, from then onwards most physicists actually used - and still
uses - the Mansouri-Sexl formalism as test-theory. Most
experimental tests of isotropy of the one-way velocity of light
are actually performed in some background inherited by the the
``Mansouri-Sexl test-theory''.

Some years later, the Mansouri-Sexl gauge was extended by
Anderson, Vetharaniam and Stedman (see \cite{avs} and references
therein) to such an extent that the one-way velocity of light
became even more conventional, provided that it complies with a
suitable synchronization gauge. Of course we are going to name
this gauge, which is very wide indeed, ``AVS synchronization
gauge'' or briefly ``AVS-gauge''. Recently, simultaneity and
synchronization gauges have been studied in a  very interesting
paper by Minguzzi\cite{minguzzi03}.

\subsection{Between conventionalism and realism: Selleri's approach}

In recent years, Franco Selleri
\cite{selleri4,selleri3,selleri,selleri2} has sided against both
the conventionalist approach and the SRT, maintaining the cause of
realism in a Lorentz-like background; yet he agrees with one of
the main points of the conventionalist attitude, namely the
statement that, in any IRF, ``the costancy of the one-way speed of
light is a mere convention without any empirical
cornerstone''\footnote{Translation  by the authors.}. Unlike the
AVS approach, Selleri approach is not formal but truly physical,
or to be exact conceptual, and even philosophical; it is not
surprising that such an approach gave rise to a wide and
fascinating discussion about the foundations of SRT
\cite{selleri4,selleri3,selleri,selleri2}, \cite{bergia},
\cite{bergia-guidone}, \cite{bergia-valleriani}, \cite{RS}.

In particular, Selleri has undertaken a severe critical analysis of the two
principles of SRT (relativity principle and invariance of the velocity of
light), emphasizing some aspects which, after a century from their first
enunciation (1905), seem to be still somehow `blurred': both concerning
their formulation in precise operational terms and concerning their
connection to precise empirical data.

Selleri bases his reconstruction of theoretical physics upon three ``hard
experimental evidences'', on which the whole scientific community is in full
agreement. They are, as Selleri himself stresses, hypotheses directly
supported by the experimental data and, thus, wholly independent on any
theoretical license. At variance with the two principles of SRT that, being
so loaded with theory, were formulated, in the celebrated 1905 paper,
without any reference to experimental data.\footnote{%
Actually, as well known, Einstein's aim was basically to recover the `unity
of physics' which, at the beginning of the twentieth century, seemed to have
been lost because of the existence of two distinct invariance groups: one
pertaining to Newton's mechanics and the other one valid for
Maxwell-Lorentz's electromagnetism. Einstein thus aimed at rebuilding the
theoretical physics so that all its branches share for the same invariance
group.}

Suspicious of any preconceived theoretical framework, and careful only at
the experimental data, Selleri shows that there exist not only a theory, but
a whole set of theories all compliant with the `hard experimental
evidences', which can be distinguished between each other by the value of a
parameter $e_{1}$ (called ``synchronization parameter''), which is \emph{\`a
priori} unconstrained. However, Selleri's realistic attitude cannot be
satisfied with this result, which apparently supports an unwanted
operational viewpoint; in fact, this is only the first part of Selleri's
approach. The second part, driven by a strong realistic viewpoint, is the
attempt to prove that a secret ``Nature's synchrony choice'', although
totally hidden in the class of IRF's, actually exists, and can be unveiled
by means of suitable experiments performed in non-IRF's.

Selleri's approach can be briefly summarized by the following statements:

(i) in the ensemble of the theories consistent with the ``hard experimental
evidences'', the synchronization parameter $e_{1}$ cannot be fixed by any
experiment performed in IRF's;

(ii) the SRT belongs to this ensemble and it corresponds to a
definite  value of the parameter $e_{1}$, namely $e_{1}=-\beta
\gamma /c;$

(iii) all the theories in which $e_{1}\neq -\beta \gamma /c$ are
inconsistent with SRT;

(iv) even though statement (i) entails that the synchronization
parameter cannot be fixed by any experiment performed in
unaccelerated reference frames, such a parameter can be fixed by
suitable experiments performed in accelerated, in particular
rotating, reference frames. Selleri \cite{selleri} actually,
considering some of such experiments (primarily the Sagnac
effect), finds that Nature forces the synchrony choice $e_{1}=0$,
which is different from the relativistic one $e_{1}=-\beta \gamma
/c$.

The choice $e_{1}=0$ turns out to be consistent with the
relativity of space and time, but not with the relativity of
simultaneity. As a consequence ``absolute simultaneity'' can be
re-introduced in physics, against the ``relative simultaneity''
(which is a typical feature of SRT) and in agreement with the
``realistic'' (Lorentz-like) ideological assumption that a
privileged IRF, namely the IRF in which the ether is at rest,
actually exists.

\subsection{Plan of the paper}

For sake of clarity, we split this paper into two parts. In the
first one, far and away the most extensive
(Secs.~\ref{sec.thesis}-\ref{sec.S. gauge in full SRT}), only
IRF's will taken into account; in the second part
(Sec.~\ref{sec.rotating ref. frames}) we will briefly consider
also rotating reference frames. It is ultimately in this second
part that a disagreement with Selleri's approach and conclusions
will emerge: in particular, conclusion (iv) will be disproved. But
all along the first part, i.e. until accelerations are neglected,
a basic agreement on Selleri's conclusions, in particular
conclusions (i), (ii), will appear. Let us emphasize that we shall
give a mathematical proof of Selleri's basic idea, namely
statement (i). Actually, such a statement is a mere conjecture,
expressed in Ref. \cite{selleri3} in the following ``weak'' form:
\textit{in
the class of IRFs, no physical experiment can discriminate the case}\ $%
e_{1}=0$\ \textit{from the case}\ $e_{1}=-\beta \gamma /c$. As a matter of
fact, Selleri does not prove this statement, but only disproves specific
attempts to discriminate the \textit{case}\ $e_{1}=0$\ from the case\ $%
e_{1}=-\beta \gamma /c$ by means of some empirical evidence. Selleri
concludes: ''I'm convinced that with a bit of work this theorem can be
[proved and] extended to the set of all possible values of $e_{1}$''%
\footnote{Translation  by the authors.}.

A masterly suggestion. This theorem will be proved for the set of all
possible values of $e_{1}$ even twice: first in Sec.~\ref{sec.synthesis}
following Selleri's approach and showing its compatibility with SRT; then in
Sec.~\ref{sec.S. gauge in full SRT} in the full theoretical background of
SRT.\bigskip

Until accelerated frames will be neglected, the only thing we do not agree
is statement (iii): however, we strongly suspect that this disagreement
simply depends on an improper use of the expression ``SRT'', which is, all
over relativistic literature, commonly identified with the expression
``standard formulation of the SRT''. In our opinion, until the SRT is
identified with its standard formulation Selleri is perfectly right, since
statement (iii) is an unavoidable conseguence of the standard formulation of
the axiomatic basis of SRT. As a consequence, the incorrect statement (iii)
imperiously shows the need of recasting the axiomatic basis of SRT in a more
appropriate way, overriding - once and for all - the ambiguities that have
been passed down over almost a century. This way Selleri's approach turns
out to be very useful towards a deeper understanding of SRT, not only with
regard to the points (i), (ii) with which we agree, but also with regard to
the point (iii), with which we don't agree.\bigskip

In this paper the axiomatic basis of SRT will be re-examined in a
very thorough way from an operational point of view, with
particular emphasis on some crucial details concerning the status
of Einstein synchronization (Sec. \ref{sec.post. of SRT rev}). In
this reassessment, a central role will be played by the so-called
`round-trip axiom' (Sec.~\ref{ssec.round-trip}): the only possible
`synchrony-independent' formulation, strongly supported by
empirical evidence, of the principle of invariance of velocity of
light. We will then show (Sec.~\ref{sec.E/S reconcilied}) that,
once correctly and explicitly phrased, the principles of SRT are,
at the kinematical level, fully equivalent to the three 'hard
experimental evidences': tuning the values of the parameter
$e_{1}$ with respect to the one corresponding to SRT, one gets
theories which, at variance with Selleri's claims, are not at all
alternative to SRT, but constitute a simple rewriting of Special
Relativity. In other words, one obtains `theories' that differ
from SRT for the difference in the chosen synchronization
procedures but that are wholly equivalent to SRT in predicting
empirical facts. Technically speaking, this will result in the
introduction, in the full background of SRT, of what we are going
to name ``Selleri synchronization gauge'' (Sec.~\ref
{sec.antithesis}). Such a gauge, which turns out to be physically
meaningful (unlike the AVS-gauge, wider but rather formal) and
leaving all the observables unchanged, enables us to face the
relationships between Einstein's relativity and the so-called
``alternative theories'' of Selleri, usually considered to be on
pretty bad terms with one other. This direct comparison will be
properly formalized, thus allowing to reconcile Selleri and
Einstein, on the ground of a careful re-examination of the
conceptual bases of SRT. In our opinion, this is a very important
contribution to the clarification of SRT, coming from Selleri's
severe critique.\bigskip

As a conclusion, we have proved that Selleri synchronization gauge
does not lead to an alternative theory with respect to SRT, but
\textit{must} be incorporated in the formalism of SRT. According
to a scholastic dialectic scheme, if SRT is the thesis and Selleri
``alternative theory'' the antithesis, a suitable reworking of
SRT, starting from the ``hard experimental evidences'' and
incorporating the Selleri synchronization gauge, should be the
synthesis. This is the scheme of the first part of the paper
(Secs.~\ref{sec.thesis}, \ref{sec.antithesis},
\ref{sec.synthesis}). The main part of the paper (Sec.~\ref{sec.S.
gauge in full SRT}) incorporates the Selleri synchronization gauge
in the theoretical background of SRT in a fully formal way, namely
as a suitable sub-gauge of the Cattaneo gauge, which is the set of
all the possible parametrization of a given physical reference
frame; in particular, Selleri's ``absolute synchronization'' -
which could sound somewhat `heretical' to `orthodox' ears -
emerges naturally from Minkowski spacetime of SRT as a simple
parametrization effect, involving a frame-invariant foliation of
spacetime.

The paper ends (Sec.~\ref{sec.rotating ref. frames}) with a brief
analysis of Selleri's statement (iv), which maintains that the
``hard experimental evidences'' should force the synchronization
parameter $e_{1}$ to take the value zero when rotating reference
frames are taken into account. Conversely
our analysis, carried out in the theoretical background of SRT \textit{%
incorporating Selleri synchronization gauge}, shows that: (i)
contrary to Selleri's claim, $e_{1}$ is a free parameter
\textit{in any case}\footnote{It could be said, partly for fun and
partly for real, that we take Selleri synchronization gauge more
seriously than Selleri himself}; (ii) the gauge freedom introduced
in SRT is not a marginal detail or a philosophical nicety, but a
very useful tool which allows a clear explanation of the Sagnac
effect, \textit{in the full background of SRT}, for
counter-propagating matter beams.

\section{Einstein's approach: postulates of SRT reviewed (thesis)}\label{sec.post. of SRT rev}
\label{sec.thesis}

In this section we aim to thoroughly analyze the operational
meaning of the two postulates of SRT, with an especial care for
their connections to Einstein synchronization procedure.

\subsection{Traditional formulation of the axiomatic basis of SRT}

As well known, the standard expression of the two postulates is the
following:

\begin{itemize}
\item[($\alpha $)]  {\bf Relativity principle}:\ \emph{all physical
laws
are the same in any IRF. No inertial reference frame is `privileged', \emph{%
i.e.~}distinguishable from the other IRF's by means of `internal' empirical
evidences.}

\item[($\beta $)]  {\bf Invariancy of the velocity of light}:
\emph{the velocity of light in empty space is the same in any IRF.
Its value is given by the universal costant }$c$\emph{.}
\end{itemize}

Let us notice that, in all the customary textbook treatments of
the theory \cite{moller, resnick, rindler}, the problems of
distant simultaneity and of the definition of the synchronization
procedure are dealt with \emph{after} the introduction of the
postulates ($\alpha $) and ($\beta $). Nonetheless, assuming that
proposition ($\beta $) retain a precise meaning without the need
of a \emph{previous} definition of a synchronization procedure can
result -- and, as we will see, does result -- into misleading
interpretations of the postulate itself. According to a strict
operational approach, it is therefore convenient to specify the
definition of Einstein synchronization procedure
\emph{independently from postulate} ($\beta $).

\subsection{Einstein synchronization\label{ssec.einstein synchr}}

Einstein synchronization procedure is defined, without any
reference to postulate ($\beta $), by the following sequence of
operations (cfr. \cite {einstein}):

\begin{enumerate}
\item  An arbitrary spatial origin of the reference frame, which will be
called $O$, is chosen. A standard clock, together with a light emitter, is
lodged in $O$. In any other point of space, to which we will generically
refer as $A$, an identical standard clock, together with a mirror, is lodged.

\item  At time $t_{0}$ the light emitter sends a pulse from $O$.
Such a pulse reaches $A$, is reflected and reaches back $O$ at
time $t_{0}+\Delta t_{0}$. Thus, if $l_{OA}$ is the spatial length
of the segment $OA$, then
the mean velocity of the light pulse along the closed trip $OAO$ is \emph{%
empirically} \emph{given} by $c_{OAO}\equiv 2l_{OA}/\Delta t_{0}$.

\item  At time $t_{1}$ a second pulse is emitted from $O$ towards $A$. At
the reception of the pulse in $A$, the clock here located is set to the time
$t_{A}\equiv t_{1}+\Delta t_{0}/2$. Thus the \emph{one-way velocity }of
light, from $O$ to $A$, is \emph{conventionally defined} by $c_{OA}\equiv
c_{OAO}=2l_{OA}/\Delta t_{0}$. In other words the set time $t_{A}$ can be
expressed in terms of the one-way velocity: $t_{A}\equiv t_{0}+l_{OA}/c_{OA}$%
.
\end{enumerate}

We mention that such a synchronization procedure, fully
conventional as it is, does not guarantee, by itself, neither the
property of \textit{optical isotropy, }nor the crucial requirement
of the \textit{transitivity} of the synchronization
procedure.\footnote{Let us point out that only \textit{\
``transitivity }is actually a vital requirement for any
self-consistent definition of simultaneity. On the contrary,
``\textit{optical isotropy''} is just a lovely feature, which - in
case - could be dropped without any problem} Only the introduction
of a proper empirical hypothesis about the propagation of light,
namely the round-trip axiom (see Sec. \ref{ssec.Operational
formulation} below), can lead to such properties. In this context
(in which the round-trip axiom is going to play a key role)
Einstein synchronization, while maintaining its conventional
character, will reveal all its theoretical usefulness and its
remarkable physical and heuristic meaning. However, we shall see
in Sec. \ref{ssec.selleri synchronization} that the widespread
statement that Einstein synchronization is the unique transitive
synchronization (that is to say the unique self-consistent
synchronization) is an untenable prejudice.

\subsection{Operational formulation of the axiomatic basis of SRT\label%
{ssec.Operational formulation}\label{ssec.round-trip}\label{ssec.roundtrip}}

We move now to investigating the link connecting postulates ($\alpha $) and (%
$\beta $) to Einstein synchronization, in order to unveil the operational
contents of the postulates. This analysis will lead to a strict formulation
of the kinematical consequences of ($\alpha $) and to an utter recasting of (%
$\beta $).\bigskip

First of all, let us translate proposition ($\alpha $) as suggested by
Bergia and Valleriani in ref. \cite{bergia-valleriani}: \textit{``the same
experiments performed under the same conditions in different inertial
systems yield the same results''}. Provided that the (rather slippery)
expression ``under the same conditions'' is properly defined, this statement
unveils the physical content of the Relativity principle. Now, let us stress
an obvious but crucial consequence of this statement at a purely kinematical
level (which we are going to name ``Kinematical Relativity principle''):

\begin{itemize}
\item[($\alpha _{1}$)]  {\bf Kinematical Relativity principle}: \emph{%
once Einstein synchronization has been performed in any IRF, the space-time
coordinate transformations between any two IRF's have to be symmetric and
dependent on the relative velocity of the two frames alone.}
\end{itemize}

This is a formal expression of the Relativity principle at kinematical level%
\footnote{%
The formal expression of the Relativity principle - in its full
generality - consists in the requirement that \textit{all the
physical laws must be covariant with respect to the group of
symmetric coordinate transformations among Einstein-synchronized
IRF's}.}, which might appear pedantic and even boring. However,
let us mention that a less specific formulation of the Relativity
principle has lead Franco Selleri to a, in our opinion
unjustified, refusal of the principle itself \footnote{%
The formulation of the principle adopted by Selleri lacks te crucial ``once
Einstein synchronization has been performed in any IRF''.}. \bigskip

Proposition ($\beta $) admits a strict operational meaning if and only if it
is interpreted as follows:

\begin{itemize}
\item[($\beta _{1}$)]  {\bf Invariancy of the velocity of light}: \emph{%
in any IRF, once Einstein synchronization has been performed, the velocity
of light is }$c$ \emph{along any path.}
\end{itemize}

The absence of an explicit reference to Einstein synchronization in the
usual formulation of the principle of invariancy of the velocity of light
has brought many authors to claim that proposition ($\beta $) cannot be
empirically tested \cite{reichenbach1, reichenbach2, mansouri, avs, selleri2}%
. While such a point of view is formally correct, we stress that the
principle, if (and only if) rigorously recast as ($\beta _{1}$), holds a
strict physical and operational meaning. Actually, an empirically testable
formulation of the principle, fully equivalent to ($\beta _{1}$) and
avoiding any reference to the synchronization procedure (a single clock
being involved), can be promptly given. Such a formulation, the so called
\emph{`round-trip axiom', }was introduced by Reichenbach \cite{reichenbach1}%
, and reads:\footnote{%
To be precise, we mention that Reichenbach's original formulation
\cite[sec. 12]{reichenbach1} slightly differs from proposition ($\beta _{2}$%
). However, in deference to the original, we keep the appellation
`round-trip axiom'.}

\begin{itemize}
\item[($\beta _{2}$)]  {\bf Round-trip axiom}: \emph{The velocity of
light is a universal constant }$c$ \emph{in any IRF along any closed path}$%
.\bigskip $
\end{itemize}

\noindent {\bf Theorem 1}\textit{:}\textbf{\ }\textit{The round-trip
axiom (}$\beta _{2}$\textit{) is equivalent to the principle of
invariancy of the velocity of light, provided it is formulated in
the operationally meaningful form (}$\beta _{1}$\textit{):}

\[
(\beta _{2})\Longleftrightarrow (\beta _{1})
\]


\begin{itemize}
\item[$\Leftarrow $]  The demonstration is immediate.

\item[$\Rightarrow $]  Let us consider an inertial frame $S$, with spatial
origin $O$, in which Einstein synchronization has been performed. Let $%
\widehat{AB}$ be a generic path (see Fig.\ref{roundtrip}) of length $l_{AB}$%
. Let a light pulse be emitted from $A$ at time $t_{1}$ and reach $B$ along $%
\widehat{AB}$ at time $t_{2}$. Our goal is showing that $t_{2}=l_{AB}/c$
exploiting ($\beta _{2}$).

Let us therefore suppose that a second pulse goes from $O$ to $A$ along the
segment $OA$, reaching $A$ at time $t_{1}$ (exactly when the first pulse is
going off), and that a third pulse goes from $B$ to $O$ along the segment $%
BO $, starting from $B$ at time $t_{2}$ (exactly when the first pulse is
coming). Let $l_{OA}$ and $l_{OB}$ be, respectively, the length of the
segments $OA$ and $OB$. Now, it is easily verified that the round-trip axiom
($\beta _{2}$) and Einstein synchronization together ensure that the
velocity of light is $c$ along any straight line passing by $O$, in both
directions.\footnote{%
It is enough to note that a two-ways trip is a round-trip, so that ($\beta
_{2}$) ensures that $c_{OBO}=c$ for any point $B$. But according to Einstein
synchronization $c_{OB}=c;$ as a consequence $c_{BO}=c$ (for any $B$)}
Therefore, the sequence of events characterizing the global space-time path
of the pulses along the closed spatial path $OABO$ is the following
\begin{equation}
\left( O,t_{1}-l_{OA}/c\right) \rightsquigarrow \left(
A,t_{1}\right) \rightsquigarrow \left( B,t_{2}\right)
\rightsquigarrow \left( O,t_{2}+l_{OB}/c\right) \mathrm{\ .}
\end{equation}

Straightforwardly applying ($\beta _{2}$) then gives
\begin{equation}
\frac{l_{OA}+l_{AB}+l_{OB}}{\frac{l_{OA}}{c}+t_{2}+\frac{l_{OB}}{c}}=c%
\mathrm{\ ,}
\end{equation}

from which one gets $t_{2}=l_{AB}/c$, which completes the proof. $\Box $
\end{itemize}

Summarizing, we have shown that ($\beta _{1}$) and ($\beta _{2}$) are two
equivalent hypotheses, each of them being empirically testable.

\section{Selleri's approach: from hard experimental evidences to
''inertial transformations'' (antithesis)\label{sec.antithesis}}

\label{dure}

After a review of the basic hypotheses of SRT according to the canonical
Einstein's approach, we want to outline here the hypotheses which Selleri's
alternative approach is based on. As we shall see, his approach is both
original and self consistent, and straightforwardly leads to a
generalization of relativistic kinematics which encompasses a multiplicity
of synchronization procedures; among them, Einstein's procedure is nothing
but a particular case.\newline

\subsection{\label{ssec:postulates1}The ``hard experimental evidences''}

Selleri \cite{selleri3} aims at finding the most general coordinate
transformations among IRFs on the ground of the following three hypotheses:%
\newline

\begin{enumerate}
\item[(i)]  \emph{\ - \ There exists at least one IRF, let us call it $S_{0}$%
, where the velocity of light is $c$ at each point and in every direction.}

\item[(ii)]  \emph{\ - \ The two ways velocity of light is the same in every
direction, in each IRF.}

\item[(iii)]  \emph{\ - \ The ticking of clocks moving with respect to $%
S_{0} $ with velocity $v$, is slowed down by the usual factor $\sqrt{1-\beta
^{2}}$ where $\beta =\frac{v}{c}$ (''retardation of clocks'').}
\end{enumerate}

These hypotheses are supported by the experimental results to such a high
degree that they appear to be practically independent of any theoretical
speculation. Indeed, Bergia \cite{bergia} considers them the ``hard core of
the experimental knowledge'' pertaining to relativistic theories: following
him, we shall call them \textit{hard experimental evidences}.

Statements (i) and (iii) requires some clarifications which take
into account the operational details outlined in Sec.
\ref{ssec.Operational formulation}.

Statement (i) asserts that in the IRF $S_{0}$, \textit{after that
an Einstein synchronization has been performed}, the velocity of
light is $c$ along any path. Notice that (i) does not rule out the
possibility of having more than one IRF where the propagation of
light is isotropic.

Statement (iii) asserts that if $\delta \tau $ is the time interval between
two events which occur at the same place in an IRF $S$, moving with a
velocity $v$ with respect to $S_{0}$, and $\delta t_{0}$ is the time
interval between the two events \textit{as measured by two clocks
Einstein-synchronized in $S_{0}$}, then
\begin{equation}
\delta \tau =\delta t_{0}\sqrt{1-\beta ^{2}}\mathrm{\ .}  \label{tempi}
\end{equation}

\noindent The asymmetry of eq. (\ref{tempi}) reflects the different
operational meaning of the time intervals $\delta \tau $ and $\delta t_{0}$.%
\footnote{$\delta \tau $ is measured by a single clock at rest in $S$; $%
\delta t_{0}$ is measured by a couple of clocks at rest in different
locations in $S_{0}$, provided that they are Einstein-synchronized in such a
frame.}

\subsection{Selleri general coordinate transformations}

That being stated, Selleri\cite{selleri} obtains the most general coordinate
transformations \textit{from the optically isotropic IRF $S_{0}$ to a
generic IRF $S$}, in motion with respect to $S_{0}$ with a dimensionless
velocity $\mathbf{\beta \equiv v/}c$, which are in agreement with (i), (ii),
(iii). If the $x,y,z$ directions of the two frames are parallel, and the
velocity $\mathbf{\beta }$ is along the $x$ direction, the coordinate
transformations turn out to be:
\begin{equation}
\left\{
\begin{tabular}{ccc}
$t$ & $=$ & $\gamma ^{-1}t_{0}+e_{1}(x_{0}-\beta ct_{0})$ \\
$x$ & $=$ & $\gamma (x_{0}-\beta ct_{0})$ \\
$y$ & $=$ & $y_{0}$ \\
$z$ & $=$ & $z_{0}$%
\end{tabular}
\right. \quad \mathrm{,}  \label{trasfo}
\end{equation}
where $\gamma \equiv 1/\sqrt{1-\beta ^{2}}$ and $e_{1}$ is an arbitrary
function of $\beta $, whose dimensions are $[velocity]^{-1}$.

As a consequence, the most general coordinate transformations consistent
with the hard experimental evidences, in the kinematical conditions outlined
above, turn out to be a family of transformations (``Selleri synchronization
gauge'') parameterized by the function $e_{1}(\beta )$. \textit{Every $%
e_{1}(\beta )$ is admissible, and each of them corresponds to a different
synchronization procedure in each IRF moving with dimensionless velocity }$%
\beta $ with respect to $S_{0}$\textit{\ (hence, $e_{1}$ is called
``synchronization parameter'')\footnote{Of course all  possible
functions \textit{$e_{1}(\beta )$ }must satisfy the limiting
condition
\[
\lim_{\beta \rightarrow 0}\mathit{e_{1}(\beta )=0}
\]
so that eqns. (\ref{trasfo}) riduce to identity for vanishing
$\beta $. Provided this obvious constraint is satisfied, the
\textit{\ }IRF $S_{0}$ turns out to be Einstein-synchronized for
any choice of the synchronization parameter \textit{$e_{1}$}.}
and, according to Selleri, ``a different
theory''.}\footnote{%
The latter terminology expresses Selleri's conviction that the
synchronization parameter can be determined on the basis of the
observational results in accelerated frames. This means that, in
Selleri's opinion, the ``hard experimental evidences'' should lead
to a test-theory, provided that accelerated reference frames are
taken into account.}\bigskip

From (\ref{trasfo}) it is possible to obtain the velocity of light in $S$ as
a function of the angle $\vartheta $ between the propagation direction and $%
\mathbf{\beta }$:
\begin{equation}
\widetilde{c}(\vartheta )=\frac{c}{1+\Gamma \cos \vartheta }\mathrm{\qquad ,}
\label{velocità2}
\end{equation}
where
\begin{equation}
\Gamma \equiv \beta +{e_{1}(\beta )}c\gamma ^{-1}  \label{gamma}
\end{equation}
is a function of the of the dimensionless velocity $\beta $ (in
absolute value) and depends on the synchronization parameter
$e_{1}$, i.e. on the synchrony choice inside the Selleri
gauge.\footnote{Notice that $\Gamma $, as well as $e_{1}(\beta )$,
is independent of the space-time coordinates of the IRF, for each
synchrony choice.}\bigskip

\textbf{Does Selleri synchronization gauge challenge the
Relativity principle? \medskip }

As extensively pointed out by Selleri, \textit{when $\Gamma =0$, i.e. when $%
e_{1}(\beta )=-\beta \gamma /c$, each IRF becomes optically isotropic, and (%
\ref{trasfo}) reduces to the special Lorentz transformation along
the $x$ axis.} In this synchrony choice (``Einstein synchrony
choice''), and only in this synchrony choice, the coordinate
transformations (\ref{trasfo}) take a lovely symmetric form,
clearly reflecting the Relativity principle into the formalism.
But how about non-Einstein synchrony choices (\textit{$e_{1}(\beta
)\neq -\beta \gamma /c$)}?

The widespread but, as we have seen, ambiguous formulation ($\alpha $) of
the Relativity principle leads Selleri to utterly reject the Relativity
principle for any value of $e_{1}$ different from \textit{$-\beta \gamma /c$}%
. In any non-Einstein synchrony choices, the asymmetry of transformations (%
\ref{trasfo}), relating the optically isotropic IRF $S_{0}$ to the optically
anisotropic IRF $S$, leads Selleri to the following claim \cite{selleri}:
``transformations (\ref{trasfo}) violate the Relativity principle for any $%
e_{1}$, except for the relativistic value \textit{$-\beta \gamma
/c$}''; and coherently concludes: ``Varying $e_{1}$ one obtains
different theories, all equivalent to SRT as far as the
explanation of the most known experimental results is concerned.
However, any theory different from SRT is not compatible with the
principle of Relativity.''\footnote{Translation by the
authors.}\bigskip

\label{ssec:postulates2}This critical attitude towards the Relativity
principle is based, as we have already clarified, on a wrong formulation of
the principle on the kinematical level. In fact one cannot require the
symmetry of the transformations without imposing the fundamental condition
\emph{``once the same synchronization procedure has been adopted in any IRF''%
}. The crucial point here is that, in the IRF $S$, the synchronization
procedures resulting from values of $e_{1}$ different from the relativistic
one depend on the relative velocity $\mathbf{\beta }$ of $S$ with respect to
the `privileged' (\textit{i.e.~}Einstein-synchronized) IRF $S_{0}$.
Therefore the resulting transformations' asymmetry, far from violating the
Relativity principle, simply reflects into the formalism the difference
between the synchronization procedures adopted in $S$ and $S_{0}$.

In other words, \textit{the asymmetry of the transformations (\ref{trasfo})
does not accord any kind of privilege, on the physical ground, to the IRF }$%
S_{0}$\textit{\ in which the adoption of Einstein synchronization has been
stipulated, but constitutes the formal expression of the asymmetry of the
synchronization procedures adopted in }$S$\textit{\ and }$S_{0}$\textit{\
respectively}.\textit{\ }

On the other hand, it would be a striking surprise if, adopting in $S$ a
non-Einstein synchronization procedure, depending on $\mathbf{\beta }$ by
definition, whereas an Einstein synchronization procedure is adopted in $%
S_{0}$, we could obtain transformations relating the two frames
that are symmetric and not depending on $\mathbf{\beta }$!\bigskip

Summarizing, the `privileged role' played by $S_{0}$ in Selleri's theory is
a mere artificial element, $S_{0}$ being just the IRF in which, \emph{by
stipulation}, Einstein synchronization has been performed: as a matter of
fact, \emph{any IRF }$S$\emph{\ can play the role of }$S_{0}$.

\subsection{Selleri synchronization\label{ssec.selleri synchronization}}

In operational terms, each synchronization procedure belonging to Selleri
gauge can be obtained, in the inertial frame $S$ moving with respect to $%
S_{0}$ with (dimensionless) velocity $\mathbf{\beta }$, by means of the
following operations, at each point $A\in S$:

\begin{itemize}
\item[{a}]  First of all, let us choose an arbitrary origin of the
spatial coordinates, that we shall call $O$. Let us suppose that a
standard clock and a source of light signals are lodged in $O$.
Identical clocks and mirrors are lodged in all points of space;
let $A$ be one of these points.

\item[{b}]  At $t_{o}$ the light source in $O$ emits a signal.
This signal reaches $A$, is reflected by the mirror and comes back
to $O$, where it arrives at $t_{o}+\Delta t_{o}$. Consequently, if
$l_{\left( OA\right) }$ is the spatial length of $OA$, the mean
velocity of the signal along the closed path $OAO$ \textit{is}
$c_{\left( OAO\right) }\equiv 2l_{(OA)}/\Delta
t_{o}$.\footnote{%
This is an \textit{empirical} velocity.}

\item[{c}]  At $t_{1}$ another signal propagating to $A$ is
emitted. Let $\vartheta _{OA}$ be the angle between $\mathbf{\beta
}$ and $OA$; when it receives this signal, the clock lodged in $A$
is set in such a way to read the time $t_{A}\equiv
t_{1}+l_{(OA)}/\widetilde{c}(\vartheta _{OA})$, where
$\widetilde{c}(\vartheta _{OA})$ is the one way velocity of the
signal
(from $O$ to $A$) \textit{defined by} (\ref{velocità2}):\footnote{%
This is a \textit{conventional} velocity, because the
dimensionless parameter $\Gamma $ is an arbitrary function of
$\beta $.}
\begin{equation}
t_{A}\equiv t_{1}+l_{(OA)}/\widetilde{c}(\vartheta
_{OA})=t_{1}+l_{(OA)}\left( 1+\Gamma \cos \vartheta _{OA}\right)
/c
\end{equation}
An obvious consequence is that, in general, $c_{(OA)}\neq
c_{\left( OAO\right) }$: this expresses the fact that a generic
inertial frame $S$, different from $S_{0}$, is not optically
isotropic for any value of $\Gamma $ different from the
relativistic value $\Gamma =0$.
\end{itemize}

\noindent As we shall show in Sec. \ref{sec.E/S reconcilied}
({Theorem 3}), the velocity of light $\widetilde{c}(\vartheta
_{OA})$ given by (\ref{velocità2}) implies that the observable
quantity ``time of flight of a light pulse along a generic closed
path'' must be in agreement with the
round-trip axiom for any synchrony choice, i.e. for each value of\textit{\ $%
\Gamma $.\footnote{In particular, the observable quantity $\Delta
t_{o}$, expressing the time of flight of the signal along the
there and back path $OAO$, turns out to be in agreement with the
round-trip axiom \textit{for each value of $\Gamma $}}.}
\medskip

That having been said, the crucial question is the self-consistency, namely
the \textit{transitivity}, of Selleri synchronizations. This is not a matter
of conventions but a matter of facts, because the transitivity of a
synchronization procedure is a fact empirically testable, just like the time
of flight of a light pulse along a closed path. \medskip

\noindent {\bf Theorem 2: }\textit{the transitivity of Selleri
synchronization procedures, for any value of the synchronization parameter }$%
e_{1}(\beta )$\textit{, is fully equivalent to the round-trip axiom.\medskip
}

The proof of this important theorem takes almost one page; so it
is boxed up in Appendix.

\subsection{\label{ssec:postulates3} Selleri ``inertial transformations''%
\label{ssec.Selleri in. tr.}}

According to Selleri\cite{selleri2}, the ``gauge choice''
$e_{1}=0$ is the only one that allows to get rid of the
inconsistencies between the weakly accelerated systems and the
inertial frames: we shall go deeper into the details of such a
matter below, in Sec. \ref{sec.rotating ref. frames}. Let us just
point out here that all  gauge choices are equally legitimate,
including of course $e_{1}=0$.\newline

If we set $e_{1}=0$ in (\ref{trasfo}), we obtain the following ``\textit{%
inertial transformations''} (according to Selleri's terminology):
\begin{equation}
\left\{
\begin{tabular}{ccc}
$t$ & $=$ & $\gamma ^{-1}t_{0}$ \\
$x$ & $=$ & $\gamma (x_{0}-\beta ct_{0})$ \\
$y$ & $=$ & $y_{0}$ \\
$z$ & $=$ & $z_{0}$%
\end{tabular}
\right. \mathrm{\qquad .}  \label{inerziali}
\end{equation}

The first peculiar consequence of eqns. (\ref{inerziali}) is the \textit{%
anisotropic propagation of light} in any IRF different from $S_{0}$, in such
a synchrony choice. In fact, with this choice of $e_{1}$ the velocity of
light in $S$ becomes
\begin{equation}
\widetilde{c}(\vartheta )=\frac{c}{1+\beta \cos \vartheta }\mathrm{\qquad .}
\label{velasso}
\end{equation}

Another consequence, even more peculiar, is the occurrence of an \textit{%
``absolute synchronization''}. In fact, eq.
(\ref{inerziali})$_{I}$ expresses the relativity of time but not
the relativity of simultaneity, since $\Delta t=0\Leftrightarrow
\Delta t_{0}=0$: this means that the notion of simultaneity
between events occurring at distinct points of space, in this
synchrony choice, turns out to be independent of the IRF that one
considers. This could be puzzling for relativistic people,
accustomed to the relativity of synchronization: the conundrum is
that Selleri's absolute synchronization, which at first sight
could look like a weirdness in the light of a `traditional'
relativistic approach, turns out to be perfectly legitimate also
in the full context of SRT. In fact: (i) on the operational
viewpoint, it can be obtained by means of actual operations (see
Sec. \ref{ssec.selleri synchronization}); (ii) on the formal
viewpoint, in the full context of SRT, it is an unavoidable
consequence of a perfectly legitimate synchrony choice (see Sec.
\ref{sec.S. gauge in full SRT}). This issue will be further
discussed in Sec.~\ref{ssec.absolute sym in SRT}.

\section{Einstein and Selleri reconciled\label{sec.E/S
reconcilied} (synthesis)\label{sec.synthesis}}

It is generally given for granted, both in Selleri's papers and in the ones
of Selleri's opponents, that the ``hard experimental evidences'', from which
the general coordinate transformations (\ref{trasfo}) follow, are more
general than the postulates of SRT. More explicitly, it is widely given for
granted that the postulates of SRT directly imply Lorentz transformations,
which in turn emerge from the general coordinate transformation (\ref{trasfo}%
) when \textit{$e_{1}(\beta )=-\beta \gamma /c,$ }namely in the
Einstein synchrony choice: the only one allowed by the postulates.
Contrary to this widespread belief, we are going to prove a
theorem which shows that the hard experimental evidences are not
more general than the postulates of SRT but are completely
equivalent to them, provided that the operational formulation of
the postulates of SRT (see Sec. \ref{ssec.Operational
formulation}) is adopted. \bigskip

\noindent {\bf Theorem 3. }\textit{The three ``hard experimental
evidences'' (i)-(iii), on which Selleri's theory rests, are
equivalent to postulates (}$\alpha _{1}$\textit{)-(}$\beta
_{1}$\textit{) of Einstein's SRT: }
\[
(i)\wedge (ii)\wedge (iii)\Longleftrightarrow (\alpha _{1})\wedge (\beta
_{1})\Longleftrightarrow (\alpha _{1})\wedge (\beta _{2})\mathrm{\ .}
\]

Proof.

$\Leftarrow $ The demonstration of this direction is immediate: it
is enough to notice that $(i)$, $(ii)$ and $(iii)$ are manifestly
compliant with SRT.

$\Rightarrow $ Consider a generic IRF $S$ moving with dimensionless velocity
$\mathbf{\beta }$ with respect to $S_{0}$. Let $\sigma $ be a generic closed
spatial path, at rest in $S$, $dl$ the line element of $\sigma $, $\Delta
l=\oint_{\sigma }|\mathrm{d}l|$ the length of $\sigma $ and $\Delta t$ the
time interval, measured by a single clock at rest on $S$, taken by the light
to perform a round trip of $\sigma $. The observable quantity $\Delta t$ is
given by
\begin{equation}
\Delta t=\oint_{\sigma }\frac{|\mathrm{d}l|}{\widetilde{c}(\vartheta )}%
=\oint_{\sigma }\frac{|\mathrm{d}l|}{c}+\oint_{\sigma }\frac{\Gamma \cos
\vartheta }{c}|\mathrm{d}l|\mathrm{\ ,}  \label{tempototale}
\end{equation}
where eq.(\ref{velocità2}) has been applied. The last term of eq. (\ref
{tempototale}) is the line integral of a constant vector field on a closed
path, and obviously vanishes. Therefore
\begin{equation}
\Delta t=\oint_{\sigma }\frac{|\mathrm{d}l|}{c}=\frac{\Delta l}{c}\mathrm{\ .%
}  \label{tempototale1}
\end{equation}

Eq.(\ref{tempototale1}) clearly shows that \textit{any velocity of
light of the form (\ref{velocità2}) - whatever the choice of the
synchronization parameter }$e_{1}$\textit{\ - complies with the
round-trip axiom (}$\beta _{2}$\textit{).} This is an alternative
form of {Theorem 3.}

On the other hand, we have shown in Sec. \ref{ssec.Operational formulation} (%
{Theorem 1}) that  the round-trip axiom ($\beta _{2}$) is
equivalent to the principle of invariancy of the velocity of
light, provided that it is formulated in the operationally
meaningful form ($\beta _{1}$); briefly, $(\beta
_{2})\Longleftrightarrow (\beta _{1})$. As a consequence, the
validity of the \textit{observable} relation (\ref{tempototale1})
legitimates the adoption of Einstein synchronization not only in
the ''formally privileged'' IRF $S_{0}$, but also in the generic
IRF $S$; that is to say \textit{in any IRF}. The adoption of such
a synchronization in any IRF can thus be seen as a useful
convention, fully allowed by the observative data expressed by the
round-trip axiom ($\beta _{2}$).

As a conclusion, we have come to the crucial conclusion that, if one
assumes, as Selleri does, the hypotheses ($i$), ($ii$) and ($iii$), then
Einstein synchronization is perfectly legitimate in any IRF, and its
adoption does not imply any loss of generality.

\textit{Once this synchronization procedure is conventionally\ adopted, any
IRF turns out to be optically isotropic\footnote{%
Recall that proposition ($\beta _{1}$) states that any IRF is isotropic with
respect to Einstein synchronization, so that $\widetilde{c}(\vartheta
)=c\Longleftrightarrow \Gamma =0$.}; then the spacetime transformations
between two IRF's get symmetric and depend only on the relative velocity
between the two frames\footnote{%
Recall also that adopting Einstein convention reduces the general
Selleri's transformation (\ref{trasfo}) to the usual Lorentz
transformation, according to the standard formulation of SRT.}.
Namely the kinematical relativity principle (}$\alpha
_{1}$\textit{) holds and }$S_{0}$\textit{\ loses its formally
privileged status.} This completes the proof. $\Box \bigskip $

The above theorem straightforwardly shows that no empirical
evidence can discriminate between different values of the
parameter $e_{1}$ of Selleri's theory. It is worth stressing that
this impossibility is fundamental and not due to accidental
reasons (like, \emph{e.g.}, limitations in experimental
technologies), being the consequence of the full equivalence
between Selleri's and SRT's axiomatic foundations.

Theories with the same axiomatic foundations are indistinguishable
by observations: they are said to be \textit{equivalent}. In our
opinion, addressing to them as to different \textit{physical}
theories is inappropriate and somewhat misleading: they look
rather as alternative formalisms of a unique physical theory. All
the possible choices of the parameter $e_{1}(\beta )$ correspond
to different formalisms of the same theory, operatively
indistinguishable from SRT, the difference among them being merely
a different choice of the synchronization procedure.\medskip

\textbf{So, how can we pick out a well-founded synchrony choice?
}\\

Summing up, each formalism ensuing from Transformations
(\ref{trasfo}) stems from a synchrony choice; we have called
\emph{Selleri synchronization gauge} the set of all possible
synchrony choices, related to different choices of the
synchronization parameter $e_{1} $. SRT, requiring the one-way
optical isotropy of all inertial systems, corresponds to a
particular gauge choice, namely \textit{$e_{1}=-\beta \gamma /c$}.
We have tried to stress that such an optical isotropy is not an
unavoidable choice forced by empirical evidences, but the combined
result of the principle ($\beta _{2}$), supported by a terrific
mass of empirical observations, and of the Einstein choice of the
synchronization procedure, which is not supported by any empirical
evidence, being fully conventional (actually, on a formal
viewpoint, it is just one of the infinite choices belonging to
Selleri gauge).

In compliance with the vast majority of the scientific community
we accord our preference to Einstein synchrony choice, since it is
definitely the simplest, the most elegant and the most fruitful
one. It allows for a drastic simplification of physical laws and
of their symmetry properties (see the Appendix, where the form of
Maxwell equations is outlined in an arbitrary synchrony choice);
it conforms to slow clocks transport; it allows for a simple
mathematical treatment of the causal structure of spacetime
(through the light-cone structure); it agrees with
Minkowski-orthogonality of 3-dimensional space with respect to the
wordlines of the test-particles of a given physical reference
frame and so on.

However, we do not agree with the standard approach to the matter of the
scientific community, who is used to assuming Einstein's choice as ``the
right one'' and Selleri's choice as ``the wrong one''; nor we agree with
Selleri's opposite approach, which simply overturns this statement. A
''right choice'', simply, does not exist.

\section{Selleri's synchronization gauge in the full context of SRT\label%
{sec.S. gauge in full SRT}}

In the previous sections we have tried to prove the compatibility
of Selleri approach with the Einstein approach of SRT starting
from  Selleri's ``hard experimental evidences''; namely, on the
formal ground, starting from the Selleri general transformations
(\ref{trasfo}). In this section we plan to get the same result, in
a quite formal way, starting from the very beginning in the full
context of SRT - and sticking to such a context in all that
follows.

In particular we shall show how: (i) any formalism belonging to Selleri
gauge may be recovered \textit{by a fully orthodox relativistic approach};
(ii) any IRF may be reparametrized in order to pass from relative
(Einstein's) simultaneity to absolute (Selleri's) simultaneity.

\subsection{Parametrizing a physical reference frame  in Minkowskian
spacetime}\label{ssec.parametrizing a PRF}

The Minkowskian spacetime of SRT is an affine pseudo-Euclidean manifold $%
\mathcal{M}^{4}$, with signature $(1,-1,-1,-1)$. \ A Physical
Reference Frame (PRF) is a time-like congruence $\Gamma $ in
$\mathcal{M}^{4}$made up by the set of world lines of the
test-particles constituting the ``reference
fluid''.\footnote{%
The concept of 'congruence' refers to a set of word lines filling the
manifold, or some part of it, smoothly, continuously and without
intersecting.} The congruence $\Gamma $ is identified by the field of unit
vectors tangent to its world lines. Briefly speaking, the congruence is the
(history of the) physical reference frame.\newline

Let $\{x^{\mu }\}=\left\{ x^{{0}},x^{1},x^{2},x^{3}\right\} $ be a system of
coordinates in a suitable neighborhood $U_{P}$ of a point $p\in \mathcal{M}%
^{4}$; these coordinates are said to be admissible (with respect to the
congruence $\Gamma $) when\footnote{%
Greek indices run from 0 to 3, Latin indices run from 1 to 3.}
\begin{equation}
g_{{00}}>0\ \ \ \ g_{ij}dx^{i}dx^{j}<0  \label{eq:admiss}
\end{equation}
Thus the coordinate lines $x^{{0}}=var$ can be seen as describing the world
lines of the $\infty ^{3}$ particles of the reference fluid, while the label
coordinates $\{x^{i}\}=\left\{ x^{1},x^{2},x^{3}\right\} $ can be seen as
the name of any particle of the reference fluid.

When a PRF has been chosen, together with a set of admissible coordinates,
the most general coordinate transformation which does not change the
physical frame (i.e. the congruence $\Gamma $) has the form \cite
{cattaneo,cattaneo-libro,moller,gron1}
\begin{equation}
\left\{
\begin{array}{l}
x^{\prime }{}^{{0}}=x^{\prime }{}^{{0}}(x^{{0}},x^{1},x^{2},x^{3}) \\
x^{\prime }{}^{i}=x^{\prime }{}^{i}(x^{1},x^{2},x^{3})
\end{array}
\right.  \label{eq:gauge_trans}
\end{equation}
with the additional condition $\partial x^{\prime 0}/\partial x^{{0}}>0$,
which ensures that the change of time parameterization does not change the
arrow of time. The coordinate transformation (\ref{eq:gauge_trans}) is said
to be `internal' to the PRF $\Gamma $ or, more simply, an \textit{internal
gauge transformation}. In particular, Eq. (\ref{eq:gauge_trans})$_{2}$ just
changes the names of the reference fluid's particles, while Eq. (\ref
{eq:gauge_trans})$_{1}$ changes the ``coordinate clocks'' of the particles.
In the following we will refer to this gauge as to the \textit{Cattaneo gauge%
}.\bigskip

\textbf{Observables as frame-dependent, but
coordinate-independent, physical quantities. }\\

Every relativist knows the famous, enlightening statement by
Minkowski: ``Henceforth space by itself, and time by itself, are
doomed to fade away into mere shadows, and only a kind of union of
the two will preserve an independent reality''. This ``kind of
union'' is, of course, Minkowsi spacetime. This is not the proper
place to face the controversial issue of the ontologic status of
Minkowsi spacetime: the only physical (or maybe metaphysical)
reality, or just a theoretical framework able to interrelate
observable events? However, this is the proper place to  point out
that, from the operational point of view, we have empirical access
only to Minkowski's ``shadows''; in fact, most \textit{observable\
}physical quantities (space, time, energy, momentum, and so on)
are frame-dependent. So, almost all of our empirical knowledge is
knowledge of shadows; but we should be quite clear about which
kind of shadows we are speaking of. ``Observables'' are not
elusive coordinate-dependent shadows of some 4-dimensional
metaphysical reality, but rather ``hard shadows'' which, according
to the physical meaning of observables, can depend on the PRF but
not on its parametrization - which is of course fully
conventional. This means that, \textit{once a PRF is given, any
observable must be gauge-invariant with respect to the gauge
transformation (\ref {eq:gauge_trans}) internal to the given PRF}.

\subsection{\label{ssec:c2}From AVS to Selleri's synchronization gauge\label%
{ssec.from AVS to S.gauge}}

Inside Cattaneo gauge, which is the set of all the possible
parameterizations of a given PRF, the transformation
\begin{equation}
\left\{
\begin{array}{l}
x^{\prime }{}^{{0}}=x^{\prime }{}^{{0}}(x^{{0}},x^{1},x^{2},x^{3}) \\
x^{\prime }{}^{i}=x{}^{i}
\end{array}
\right.  \label{eq. resincr}
\end{equation}
\emph{\ }defines a sub-gauge (the \textit{synchronization gauge})
which describes the set of all the possible synchronizations of
the PRF.\footnote{That is to say ``the set of all the possible
ways to spread time over space'' in the given PRF.}

Within the synchronization gauge (\ref{eq. resincr}), Einstein
synchronization is the only one which does not discriminate points and
directions, i.e. which is homogeneous and isotropic\footnote{%
A synchronization is called \textit{isotropic} if it implies, in a generic
IRF, a one-way speed of light which (at a given point) does not depend on
the direction.
\par
A synchronization is called \textit{homogeneous} if it implies, in a generic
IRF, a one-way speed of light which (along a given direction) does not
depend on the point\ } \cite{serafini}.

In particular, starting from Einstein synchronization, any IRF can be
resynchronized according to the following transformation:

\begin{equation}
\left\{
\begin{array}{l}
\widetilde{t}=\widetilde{t}\left( t,x^{1},x^{2},x^{3}\right) \Leftrightarrow
\widetilde{t}=\widetilde{t}\left( t,\mathbf{x}\right) \\
\widetilde{x}^{i}=x^{i}\Leftrightarrow \mathbf{\tilde{x}=x}
\end{array}
\right.  \label{eq:resincr1}
\end{equation}
where $t$ is the Einstein coordinate time of the IRF. According to eqns. (%
\ref{eq:resincr1}), the wordlines $\mathbf{x=}const$ of the test-particles
by which the IRF is made up turn out to be parametrized by the
resynchronized time $\widetilde{t}$ instead of the Einstein time $t.$

The most interesting synchronization gauge is the one in which a linear
dependence of the resynchronized time $\widetilde{t}$ with respect to the
Einstein time $t$ is imposed. In particular, Anderson, Vetharaniam and
Stedman consider, in a long series of papers which converge on the extensive
monograph \cite{avs} (1998), the synchronization gauge
\begin{equation}
\left\{
\begin{array}{l}
\widetilde{t}=t-\mathbf{\tilde{k}}\cdot \mathbf{x} \\
\mathbf{\tilde{x}=x}
\end{array}
\right.  \label{eq: avs}
\end{equation}
where $\mathbf{\tilde{k}=\tilde{k}(x)}$ is an arbitrary smooth vector field
(let us call it \textit{synchronization field}) only depending on the space
variable $\mathbf{x}$. The set of all the possible synchronizations defined
by eqns. (\ref{eq: avs}) defines the \textit{Anderson-Vetharaniam-Stedman
synchronization gauge}, the ``\textit{AVS-gauge}'' in brief. The most
interesting feature of this gauge is that it is the only synchronization
gauge which is formally consistent with the round-trip axiom. In fact, if $%
\Delta t$ is the time taken by a light signal for a generic round-trip, eq. (%
\ref{eq: avs})$_{1}$ straightforwardly shows that the resynchronized time $%
\Delta \widetilde{t}$ is again the same: $\Delta \widetilde{t}=\Delta t\,$%
.\bigskip

A non-null synchronization field $\mathbf{\tilde{k}}$ shatters the isotropy
of Einstein synchronization; moreover, any possible dependence of $\mathbf{%
\tilde{k}}$ on the space variable $\mathbf{x}$ breaks the homogeneity of
Einstein synchronization.

Since eqns. (\ref{eq: avs}) imply $\mathbf{\tilde{k}}\cdot \mathbf{x=}const$
along any wordline $\mathbf{x=}const$, the AVS-gauge can be interpreted as a
change in the origin of Einstein coordinate time for any point (or
test-particle wordline) of the IRF, generally variable from point to point
of the given IRF. The simplest particular case is $\mathbf{\tilde{k}}\cdot
\mathbf{x=}const$ everywhere in the whole IRF: in this instance, eq. (\ref
{eq: avs}) reduces to a trivial change in the origin of Einstein coordinate
time in the whole IRF.

\textit{The re-synchronization (\ref{eq: avs}) redefines the\
simultaneity hypersurfaces,} that are now described by
$\widetilde{t}=const$. Therefore, the set of these hypersurfaces
defines a foliation of spacetime which depends not only on the
IRF, as well known in Einstein synchronization (``relativity
of synchronization''), but also on the synchronization field $\mathbf{\tilde{%
k}(x)}$.\bigskip

A simple but absolutely non trivial case, namely the instance $\mathbf{%
\tilde{k}=}const$, was considered by Mansouri and Sexl \cite{mansouri} in
the 1977; as far as we know, this is the first attempt in which a non
trivial synchronization gauge is brought to attention.

Now, let us stress a very interesting case, which belongs to the
Mansouri-Sexl instance. Let $S_{0}$ be an inertial reference frame (IF) in
which an Einstein synchronization procedure is adopted by stipulation; and
let $S$ be an IRF travelling along the $x$-axis (of unit vector $\mathbf{e}%
_{1}$) with constant dimensionless velocity $\beta \equiv v/c$. If both $%
S_{0} $ and $S$ are Einstein synchronized, the standard Lorentz
transformation follows
\begin{equation}
\left\{
\begin{array}{l}
t=\gamma (t_{0}-\frac{\beta }{c}x_{0}) \\
x=\gamma (x_{0}-\beta ct_{0}) \\
y=y_{0} \\
z=z_{0}
\end{array}
\right.  \label{eq:lorentz}
\end{equation}
where $\gamma \doteq \left( 1-\beta ^{2}\right) ^{-1/2}$ is the Lorentz
factor. Now, let us re-synchronize $S$ according to transformation (\ref{eq:
avs}), in which the synchronization field $\mathbf{\tilde{k}}$ is chosen as
follows:
\begin{equation}
\mathbf{\tilde{k}\doteq -}\frac{\Gamma (\beta )}{c}\mathbf{\hat{x}}
\label{eq:k-tilde}
\end{equation}
$\mathbf{\hat{x}}$ being the unit vector in direction $x$ and
$\Gamma (\beta )$
being an arbitrary function of $\beta $.\footnote{%
Note that the synchronization field $\mathbf{\tilde{k}}$ defined by eq. (\ref
{eq:k-tilde}) does not depend on $\mathbf{x}$: this means that the case
under consideration actually belongs to the Mansouri-Sexl instance.} We get:
\begin{equation}
\left\{
\begin{array}{l}
\widetilde{t}=t+\frac{\Gamma (\beta )}{c}x=\gamma (t_{0}-\frac{\beta }{c}%
x_{0})+\frac{\Gamma (\beta )}{c}\gamma (x_{0}-\beta ct_{0}) \\
\widetilde{x}=x=\gamma (x_{0}-\beta ct_{0}) \\
\tilde{y}=y=y_{0} \\
\tilde{z}=z=z_{0}
\end{array}
\right.  \label{eq:selleri gauge1}
\end{equation}

If the function $\Gamma (\beta )$ is written as follows:
\begin{equation}
\Gamma (\beta )\equiv \beta +e_{1}(\beta )c\gamma ^{-1}  \label{eq:gamma}
\end{equation}
where $e_{1}(\beta )$ is an arbitrary function of $\beta ,$ then Eqns. (\ref
{eq:selleri gauge1}) take the form
\begin{equation}
\left\{
\begin{array}{l}
\widetilde{t}=\gamma ^{-1}t_{0}+e_{1}(\beta )(x_{0}-\beta ct_{0}) \\
\widetilde{x}=x=\gamma (x_{0}-\beta ct_{0}) \\
\tilde{y}=y=y_{0} \\
\tilde{z}=z=z_{0}
\end{array}
\right.  \label{eq:selleri gauge2}
\end{equation}
which exactly coincides with Selleri general coordinate transformations (\ref
{trasfo}).

Eqns. (\ref{eq:selleri gauge2}) define the \textit{``Selleri synchronization
gauge''}. Such a gauge can be interpreted as the set of all the possible
synchronizations of a given IRF which comply with the ``hard experimental
evidences''; in particular, it complies with the standard Einstein
expression of time dilation with respect to the IRF $S_{0}$, according to
Selleri's statement (iii).\footnote{%
Of course in the AVS-gauge, which includes the Selleri gauge, the time
dilatation - as an observable quantity - does not change; however, on the
formal point of view it takes a more complicated shape depending on the
vector field $\mathbf{\tilde{k}}$ (as well as on the dimensionless velocity $%
\mathbf{\beta }$), see Ref.~\cite{avs}.}

The arbitrary function $e_{1}(\beta )$ is nothing but Selleri's
``synchronization parameter''.

All this shows that the set of all the ``theories'' belonging to
Selleri gauge can be interpreted, \textit{in the theoretical
background of SRT}, as a set of parameterizations (in particular
synchronizations) of a given IRF. Different parameterizations give
rise only to different formalisms of the SRT, not to different
physical theories. Selleri's ``alternative theories'' are merely
``alternative writings'' of a unique physical theory, which is
nothing but the SRT. This completely agrees with the results found, \textit{%
in the theoretical background of Selleri's approach}, in Sec.~\ref
{sec.synthesis}.

\subsection{Selleri ``absolute simultaneity'' in the full context of SRT%
\label{ssec.absolute sym in SRT}}

The synchrony choice $e_{1}(\beta )\doteq -\beta \gamma /c$ (i.e.
$\Gamma (\beta )\doteq 0$) in Eqns. (\ref{eq:selleri gauge2})
gives the standard Einstein synchronization, whereas the synchrony
choice $e_{1}(\beta )\doteq 0 $ gives Selleri synchronization. In
the AVS formalism, the synchrony choice $e_{1}(\beta )\doteq 0$ is
equivalent to the choice
\begin{equation}
\mathbf{\tilde{k}\doteq -}\frac{\beta }{c}\mathbf{\hat{x}}
\label{eq.kappa}
\end{equation}
for the synchronization field $\mathbf{\tilde{k}(x)}$. In such a synchrony
choice, eqns. (\ref{eq:selleri gauge2}) take the form
\begin{equation}
\left\{
\begin{array}{l}
\widetilde{t}=\gamma ^{-1}t_{0} \\
\widetilde{x}=x=\gamma (x_{0}-\beta ct_{0}) \\
\tilde{y}=y=y_{0} \\
\tilde{z}=z=z_{0}
\end{array}
\right.  \label{inerziali1}
\end{equation}
which are nothing but the ``inertial transformations''
(\ref{inerziali}) advocated by Selleri. Now, let us consider, in
the full context of SRT, the ''puzzling consequence'' mentioned in
Sec. \ref{ssec.Selleri in. tr.} of Eqns. (\ref{inerziali1}):
``absolute simultaneity''.\bigskip

As widely pointed out by Selleri, the time transformation
expressed by (\ref {inerziali})$_{I}$ does not contain the term
which is responsible for the relativity of
simultaneity\footnote{In fact, the re-synchronization defined by
the choice $e_{1}(\beta )\doteq 0$ \ cancels out the term $-\gamma
\frac{\beta }{c}\,x_{0}$\ appearing in
Lorentz time transformation and responsible for the relativity of simultaneity%
{\textit{\ }\cite{bergia-guidone}, \cite{RS_libro}.}}; consequently, the
notion of simultaneity between ``spatially separated''{\ events}, in this
synchrony choice, turns out to be independent of the IRF that one considers.
This can be expressed by saying that the synchronization procedure described
by the transformations (\ref{inerziali})$_{I}$ is, in Selleri's terms,
``absolute''.\footnote{%
From an operational viewpoint, the absolute synchronization of an inertial
frame $S$ is obtained by setting the reading of a clock in $S$ equal to $%
\gamma ^{-1}t_{0}$ when its spatial position coincides with the
one of a clock in $S_{0}$, whose reading is $t_{0}$ (obviously,
this event is unique in the biography of a clock of $S$).}\bigskip

Absolute synchronization is interpreted by Selleri \cite{selleri5} on the
ground of a paradigm based on the actual existence of \textit{physically}
privileged IRF (at kinematical level), in agreement with the hypothesis of
the stationary ether, proper of the classical paradigm shared by Lorentz e
Poincar\'{e} (see, for instance, \cite{selleri5,lorentz}). In particular,
according to Selleri ``time is no more an infinite series of subjective
viewpoints, all of them equally legitimate, but gets a solid objectiveness,
similar to the one of pre-relativistic physics'' \cite{selleri}.

We are going to discuss the issue of ``absolute synchronization'' keeping
carefully apart (i) the actual possibility of ``absolute synchronization''
in SRT and (ii) its Lorentz-like interpretation, as outlined by Selleri.
\bigskip

{(i) }Although the expression ``absolute synchronization'' sounds
rather eccentric in a relativistic framework, we remind from Sec.
\ref {ssec.Selleri in. tr.} that, from the operational viewpoint,
it can be obtained by means of actual operations (see Sec.
\ref{ssec.selleri synchronization}). This would be enough to
legitimate the ``absolute synchronization'', but in the previous
section we have found an even more stringent justification:
\textit{from a formal viewpoint in the full context of SRT},
``absolute synchronization'' is an unavoidable consequence of a
perfectly legitimate synchrony choice. This means that the
synchronization gauge, once incorporated in the formalism of SRT,
should allow a sort of peaceful coexistence of Einstein relative
simultaneity and Selleri
``absolute'' simultaneity; to be exact, it should allow to introduce in SRT a%
{n ``absolute simultaneity'' without affecting neither the logical
structure, nor the predictions of the theory. This is possible if
the geometrical structure of Minkowski spacetime is so democratic
as to } accommodate, in its texture, both Einstein relative
simultaneity and Selleri ''absolute'' simultaneity. Is it indeed
possible?

According to the ``realistic'' Selleri's interpretation of
{absolute simultaneity,} this is simply impossible: Einstein
relative simultaneity and Selleri ``absolute'' simultaneity are in
contention, only one of them is able to fit the physical world.
This rigid viewpoint is shared by many
``orthodox'' relativists on the basis of the claim (on which we agree) that {%
the geometrical structure of Minkowski spacetime reflects the physical word
without any ambiguity; according to them, since the relativity of
simultaneity is a fundamental feature - embedded in the geometrical
structure of Minkowski spacetime - of the physical world. There is no room
for conventions about simultaneity. }

{Both viewpoints agree that one definition of simultaneity is
right and one is wrong, and they just disagree on which is right
and which is wrong.}

According to us instead, this ``peaceful coexistence'' is possible without
any difficulty because Selleri ``absolute'' simultaneity turns out to be {a
simple parametrization effect in Minkowski spacetime. Let us clarify this
claim. }

Our viewpoint is very plain: all the kinematical content of SRT is
encoded in {the geometrical structure of Minkowski spacetime,
which actually reflects the physical word, at the kinematical
level, without any ambiguity; yet the way of parameterizing the
Minkowski spacetime is indeed a matter of convention. To put it in
terms of observables: observable effects are determined uniquely
by the geometrical structure of Minkowski spacetime, which is
physically meaningful, not by the way it is parameterized, which
is physically meaningless.}

{\ }The choice $e_{1}=-\beta \gamma /c$, corresponding to Einstein
synchronization, involves a frame-dependent foliation of Minkowski
spacetime. Specifically, such a foliation is realized by the hypersurfaces
that are Minkowski-orthogonal\ with respect to the world lines of the
test-particles by which the reference frame is made up.

On the other hand, the choice $e_{1}=0$\ involves a frame-invariant
foliation, which is nothing but the Einstein foliation of the IRF $S_{0}$,
assumed to be optically isotropic \textit{by stipulation}.\bigskip

As a conclusion, since the way of foliating the Minkowski
spacetime is a matter of convention, both Einstein relative
simultaneity and Selleri ''absolute'' simultaneity can peacefully
coexist in Minkowski spacetime as different conventions, see fig.
2.


Einstein  showed the extraordinary power of explanation of the
relative simultaneity convention, as well as its heuristic
potential; as a matter of fact, the relativity of simultaneity -
provided that it is operationally well defined, and incorporated
in Minkowski spacetime as a well defined family of frame-dependent
foliations - is the deepest conceptual root of the theory, in
particular from the heuristic viewpoint.

On the other hand, Selleri has showed that, in some particular
cases, the ``absolute'' synchronization - provided that the
``privileged'' IRF $S_{0}$ is properly chosen - can actually lead
to a simpler description of facts. Let us mention some examples
proposed by Selleri: (i) the case of two twins living on two
rocket-ships moving along two congruent worldlines, shifted along
a spatial direction, who share the same experience and of course
also the same proper time, but share the same age only in a
``privileged'' IRF; (ii) the problem of a causally consistent
description of spatially separated events who get in contact
through a superluminal (tachionic) interaction; (iii) the issue of
synchronizing clocks on the Earth, by means of electromagnetic
signals travelling between the locations of the clocks via a
geostationary satellite \cite{selleri_libro}.

The point is that, while classical physics only allows for an
absolute synchronization, relativistic physics also allows for
relative simultaneity. This is enough to lead to a completely
different description of the physical world; in philosophical
terms, to a completely different paradigmatic world-view.\bigskip

{(ii) }The main\emph{\ }difference between our point of view and
Selleri's lies\emph{\ }in a different paradigmatic background.
Which,\emph{\ }in our opinion, does not properly belong to
physics, but rather to the interpretation of physics.\emph{\
}Selleri's work develops on the paradigmatic\emph{\ }background of
the ether hypothesis, which we consider as an unnecessary and
misleading superstructure - at least on the kinematical level -
that can be rejected as an ideological fossil, empty of any
operational meaning, without rejecting the possibility that an
absolute synchronization can be actually performed: this is
definitely our position.

This position does not exclude the existence of a ``local ether'', defined
as the optically isotropic IRF of the cosmic background radiation of the
region spanned by the Solar System. In fact, it is well known that such an
inertial frame exists; as a matter of fact, it is privileged for the
description of some astrophysical phenomena. What we reject, on the basis of
the Relativity principle (``equal experiments performed in the same
conditions lead to the same results''), is the idea of the existence of an
IRF where some physical laws hold, but the same laws do not hold in other
IRF's (we think, in particular, of the observable properties of the
propagation of light).

\section{Synchronization gauge and rotating reference frames}\label{sec.rotating ref. frames}

{As widely seen in the previous sections, }synchronization is a matter of
convention as far as IRF's are concerned. This can be formalized through a
suitable synchronization gauge. Starting from the so-called ``hard
experimental evidences'', Selleri suggest the synchronization gauge (\ref
{trasfo}), in which any allowed synchrony choice is fixed by the
``synchronization parameter'' $e_{1}$, and claims that such a parameter
cannot be fixed by any experiment performed in IRF's. However Selleri claims
(in our opinion inconsistently, but consistently with his own ``realistic''
Lorentz-like approach) that, as soon as rotating frames are considered,
synchronization cannot be conventional any longer: when rotation is taken
into account, the synchronization parameter $\emph{e}_{1}$\ is forced to\
take the value zero, and the ``absolute synchronization'' turns out to be
the only legitimate synchronization \cite{selleri,selleri2,selleri_libro}.
In his words, ``the famous synchronization problem is solved by Nature
itself: it is not true that the synchronization procedure can be chosen
freely, because all conventions but the absolute one lead to an unacceptable
discontinuity in the physical theory'' \cite{selleri}, \cite{selleri_libro}.

The discontinuity to which Selleri refers descends from the comparison of
the \textit{global} velocity of light on a round-trip along the rim of a
rotating platform, as measured by a single clock at rest on the rim, and the
\textit{local }velocity of light, as measured by two very near clocks at
rest in a local comoving inertial frame (LCIF) $S$ on the rim. If $R$ is the
radius of the circular rim of the platform and $\Omega $ its angular
velocity as measured in the central IRF $S_{0}$, such a discontinuity
persists even when we perform the limit $\Omega \rightarrow 0$, $%
R\rightarrow \infty $ in such a way that the peripheral velocity $\Omega R$
is kept fixed (see \cite{selleri} for details). This is the reason why
Selleri properly regards this discontinuity as an ``unacceptable'' one.

Elsewhere, see Ref. \cite{RS_libro}, we have given a simple
solution of this alleged `paradox'. Summing up, the discontinuity
is uniquely originated by the fact that, in Selleri ``absolute''
synchrony choice (we prefer to say in Einstein synchrony choice of
the central IRF $S_{0}$), the synchronization procedure is
different in the central IRF $S_{0}$ and in the LCIF $S$; to be
exact, an isotropic synchronization is chosen in $S_{0}$ and an
anisotropic synchronization is chosen, according to Eqns.
(\ref{inerziali}), (\ref {velasso}), in any LCIF on the rim. This
means that the discontinuity found by Selleri is not a physical
discontinuity, but merely expresses different synchrony choices in
the involved IRF's: as a matter of fact, this discontinuity
disappears if the same synchronization procedure is adopted in any
(local or global) IRF, according to a proper formulation of the
Relativity principle (see Sec. \ref{ssec.Operational
formulation}).

Let us stress that the issue which leads Selleri to mistake
different synchrony choices in different IRF's for an
``unacceptable physical discontinuity'' is nothing but the
standard formulation of the Relativity principle; as a matter of
fact, the optical isotropy of an IRF is misleadingly considered,
in the standard textbooks of SRT, as a physical property of the
IRF itself, rather than a combined consequence of observable
physical properties of light propagation (the round-trip axiom)
and of a ``suitable'' (conventional) synchrony gauge choice in the
considered IRF. Towards the discontinuity found by Selleri we have
just two possibility: rejecting the SRT, as Selleri does, or
reformulating the Relativity principle, as we have done in Sec.
\ref{ssec.Operational formulation}.

In our opinion, Selleri's discontinuity is a lightening tool,
which unveils the inadequacy not of SRT, but of the standard
formulation of SRT: a noteworthy contribution to the understanding
of the theory, coming from an alleged antirelativist.

In particular, disagreeing both with Selleri and with many
relativistic authors, who claim that the choice of the
synchronization parameter $e_{1}$ is forced by suitable empirical
evidences (just disagreeing on the actual value of this
parameter), we claim that no empirical evidence is able to force
the choice of $e_{1}$. Neither the GPS evidence (see below), nor
the empirical evidence of Sagnac interference fringes of two light
or matter beams counter-propagating along the rim of the platform
\cite{RR_libro}, nor the empirical evidence of the difference in
the ages of two slow travelling twins, after a complete round-trip
in opposite directions \cite{RS_libro}, nor any other experimental
evidence. For an thorough, critical discussion of synchronization
problems in rotating (and, more generally `non time-orthogonal')
reference frames we cross-refer to Ref. \cite{libro}.

Yet, we are going to conclude this section facing two experimental
evidences, performed in a rotating reference frame, in which both the
Einstein and the Selleri synchrony choices can be profitably used and
confronted.

\subsection{A case where Selleri synchrony choice is fitting: a simplified
description of the GPS working}

In this section we shall limit ourselves to consider a very
peculiar rotating reference frame: the Earth. The aim of this
section is to account for the good working of the global
positioning system (GPS) in Selleri gauge in a suitable IRF,
namely the IRF $S_{T}$ in which the terrestrial axis is at rest
(called by Ashby \cite{ashby_libro} ``Earth-Centered IRF''{)}.
Notice that the good working of the GPS is indeed the most
relevant testground for relativistic kinematical (as well as
gravitational) effects in rotating systems \cite{ashby_libro}. But
is it, in particular, a testground of isotropic propagation of
light, as often claimed by several authors?

It should be clear from the previous sections that we have direct
empirical access only to the roundtrip speed of light, not to the
one-way speed of light; in particular, we know from Sec.
\ref{ssec.parametrizing a PRF} that, in a given PRF, any
observable must be invariant with respect to any synchronization
gauge\footnote{Which is a sub-gauge of the gauge transformation
(\ref{eq:gauge_trans}) internal to the PRF.}. Such a matter should
be thus closed. However, we cannot simply ignore so many
authoritative opposite opinions, often supported by lots of
(alleged) experimental evidences \cite{van flandern},
\cite{bergia}, \cite{bruni}. For instance, Van Flandern \cite{van
flandern} claims that ``the system has shown that the speed of
radio signals (identical to the 'speed of light') is the same from
all satellites to all ground stations at all times of day and in
all directions to within $\pm 12$ meters per second (m/s). The
same numerical value for the speed of light works equally well at
any season of the year''; ``our result here merely points out that
the \textit{measured} speed [of light] does not change as a
function of time of day or direction of the satellite in its orbit
when the clock synchronization correction is kept unchanged over
one day''. This extract is cited by Bergia \cite{bergia},
\cite{bergia-valleriani} who comments: ``it is hard to think that
the intricate network of \textit{cross controls} would not cause
anomalies if the isotropy were not a real
data"\footnote{Translation  by the authors.}, and concludes:
``very compelling limits on a parameter related to $e_{1}$ were
obtained by \cite{bruni}''. This means that Selleri gauge is
considered by many authors as a test-theory, and the GPS was used
as a tool aimed at determining, within compelling limits, the
synchronization parameter $e_{1}$. The result is a strong evidence
for isotropic propagation of light, basically ``because the system
is too effective. In other words, the localization based on the
hypothesis that the speed of light is independent on the direction
is too accurate. Something that sounds like a verification - very
accurate indeed - of such a hypothesis''.\footnote{Translation
 by the authors.}

Facing with so many claims which support the idea that the (terrifically
good) working of the GPS should be considered as an empirical evidence for
the isotropic propagation of light in the Earth-Centered IRF $S_{T}$, we are
forced to study this issue in detail.\bigskip


\textbf{Preliminary approach in Einstein synchrony choice.}\\ First
of all, let us explain how the GPS works by means of a basic
bidimensional example and relying on Einstein synchronization in a
suitable IRF $S_{T}$ (see below), as a conventional but useful
framework (this is what Selleri would call ``absolute
synchronization''). Recall that the aim is to determine the
position $\mathbf{r}_{I}$ of an object on which a GPS device $I$
is installed. The device $I$ disposes of a clock which has not yet
been synchronized with other clocks. Let us consider the IRF
$S_{T}$ in which the
terrestrial axis is at rest and let us synchronize the GPS satellites $A$, $%
B $ and $C$ according to Einstein procedure in $S_{T}$ (starting from the
`central station' $O$ fixed in $S_{T}$). At a given time of its clock,%
\footnote{We assume, for simplicity, a simultaneous reception of
the signals. This results in more simple expressions, without any
loss of generality.} $I$ receives from the three satellites $A$,
$B$ and $C$ the following information: the emission times $t_{A}$,
$t_{B}$ and $t_{C}$ of the signals and the positions
$\mathbf{r}_{A}$, $\mathbf{r}_{B}$ and $\mathbf{r}_{C}$ of $A$,
$B$ and $C$ at the times $t_{A}$, $t_{B}$ and $t_{C}$
respectively. For simplicity, let us assume that $O$, $I$, $A$,
$B$ and $C$ belong to the same plane. The position
$\mathbf{r}_{I}=(x_{I},y_{I})$ is then determined by the
following system, in terms of the cartesian coordinate $x$, $y$ of the plane{%
\
\begin{equation}
\left\{
\begin{array}{lll}
\sqrt{(x_{I}-x_{A})^{2}+(y_{I}-y_{A})^{2}} & = & c(t_{I}-t_{A}) \\
\sqrt{(x_{I}-x_{B})^{2}+(y_{I}-y_{B})^{2}} & = & c(t_{I}-t_{B}) \\
\sqrt{(x_{I}-x_{C})^{2}+(y_{I}-y_{C})^{2}} & = & c(t_{I}-t_{C})
\end{array}
\right. \mathrm{,}  \label{sist}
\end{equation}
}whose solution corresponds to the intersection point of three
circumferences.{\ The three dimensional extension obviously requires another
satellite. Notice however that actual GPS devices exploit the data of about
ten satellites at a time. }

If the clock in $I$ had already been Einstein synchronized (in $S_{T}$) with
the other clocks, then the system (\ref{sist}) would be `overdetermined'.
But this is not the case, since $t_{I}$ is an additional unknown of the
system (apart from $x_{I}$ and $y_{I}$). Now, obtaining $t_{I}$ as a
solution of system (\ref{sist}) is equivalent to stipulate that the one-way
velocity of light between any one of the satellites and $I$ takes the value $%
c$. In other words, the clock in $I$ is so implicitly Einstein synchronized
with all the other clocks of $S_{T}$. This can be rephrased stating that one
of the three equation of system (\ref{sist}) synchronizes \textit{\`{a} la
Einstein} the clock in $I$, while the other two determine the position $%
\mathbf{r}_{I}$ as the intersection of two circumferences.\footnote{%
The other intersection, due to the quadraticity of the system, is
neglected. In practice the employ of several satellites make the
intersection univocal.} Such circumferences represent the points
that the two signal reaches ``simultaneously'' at the instant
$t_{I}$ (where ``simultaneously''
means {\em ``simultaneously for Einstein synchronization''}).\footnote{%
Actually the synchronization refers to clocks at rest in{\ }${S}_{T}$%
\thinspace . However the clock of $I$ moves with dimensionless velocity $%
\beta _{I}$ with respect to{\ }${S}_{T}$\thinspace and gets thus
desynchronized by a factor $\gamma ^{-1}\equiv \sqrt{1-(\beta _{I})^{2}}$;
the same remark applies to the satellites' clocks. This makes no difference
on the conceptual level. The time dilations effects are automatically
corrected by the GPS software (cfr., \textit{e.g.},Ref.~\cite{ashby}).
Another correction implemented by the GPS software is needed because of the
gravitational redshift, experienced by the satellites which are placed at
different heights in the Earth's potential. These corrections will be
understood in the following.}\bigskip

\textbf{General approach in Selleri gauge. } \\

We now aim  at generalizing the previous argument, showing that
the effectiveness of the GPS can be accounted for on the basis of
any synchronization procedure belonging to Selleri gauge. We
recall that {the adoption of a synchronization procedure defined
by a certain choice of }$e_{1}(\beta )$ is equivalent to the
reparameterization of the IRF (in this instance of $S_{T}$) by the
resynchronization (\ref{eq: avs}) with the choice
(\ref{eq:k-tilde}). {\ Let $x_{I}$ , $y_{I}$ and $t_{I}$ be the
solutions of the system (\ref{sist}), that is to say the values of
space and time provided by the GPS in Einstein synchrony choice
(}$\Gamma =0${). To explain the proper working of the GPS under
any choice of }$e_{1}${\ one has to show the following
theorem.\bigskip }

\noindent{\bf Theorem 4.} \textit{Let }$t_{I}$\textit{,
}$x_{I}$\textit{, }$y_{I}$\textit{\ the coordinates of an event in
the Earth-Centered Einstein synchronized IRF
}$S_{T}$,\textit{solutions of the system (\ref{sist}); and
let }$\widetilde{t}_{I}$\textit{,\ }$\widetilde{x}_{I}$\textit{, }$%
\widetilde{y}_{I}$\textit{\ the resynchronized coordinates defined, in
agreement with Eqs. (\ref{eq: avs}), by \ }
\begin{eqnarray}
\widetilde{t}_{I} &=&t_{I}+\frac{\Gamma }{c}x_{I}  \nonumber \\
\widetilde{x}_{I} &=&x_{I}  \label{sol} \\
\widetilde{y}_{I} &=&y_{I}  \nonumber
\end{eqnarray}

\noindent\textit{Then the coordinates } $\widetilde{t}_{I}$\textit{,\ }$%
\widetilde{x}_{I}$\textit{, }$\widetilde{y}_{I}$\textit{\ are solutions of
the system (\ref{sist}) recast in the resynchronized chart }$\left\{ \tilde{t%
},\widetilde{x},\tilde{y},\tilde{z}\right\} $\textit{. In particular, the
spatial position of the device }$I$\textit{\ is the same as in the Einstein
chart }$\left\{ t,x,y,z\right\} $\textit{. }

\textit{\smallskip }

\emph{Proof.}{\ We aim at rewriting Eqs. (\ref{sist}) in the
resynchronized chart $\left\{
\tilde{t},\widetilde{x},\tilde{y},\tilde{z}\right\} $. Since the
instant }$t_{A}$ of Einstein chart corresponds, in the
resynchronized chart, to the instant {\
\begin{equation}
\widetilde{t}_{A}=t_{A}+\frac{\Gamma }{c}x_{A}\mathrm{\ } {\quad ,}
\label{pollo}
\end{equation}
the time interval $\widetilde{t}_{I}-\widetilde{t}_{A}$ needed by the signal
to go from }$A$ to $I${, over a space distance }$l_{(AI)}${, is given -
following Eq. (\ref{velocità2}) - by }

{\
\begin{eqnarray}
\widetilde{t}_{I}-\widetilde{t}_{A} &=&\int_{A}^{I}\frac{\left| \mathrm{d}%
\mathbf{l}\right| }{\widetilde{c}(\vartheta )}=\int_{A}^{I}\frac{\left|
\mathrm{d}\mathbf{l}\right| }{c}+\int_{A}^{I}\frac{\Gamma \cos \vartheta }{c}%
\left| \mathrm{d}\mathbf{l}\right| =\frac{l_{(AI)}}{c}+\Gamma \int_{A}^{I}%
\frac{\mathrm{d}x}{c}\mathbf{=} \nonumber \\
&=&\frac{l_{(AI)}}{c}+\Gamma \frac{(x_{I}-x_{A})}{c}\mathrm{\qquad .}
\end{eqnarray}
}where $\vartheta $ is the angle between $\mathbf{\beta }$ and $AI$ while $%
\cos \vartheta |\mathrm{d}\mathbf{l|=}\mathrm{d}x$. {Analogous expressions
hold for the time intervals }$\widetilde{t}_{I}-\widetilde{t}_{B}$ and $%
\widetilde{t}_{I}-\widetilde{t}_{C}$ {needed by the signal to
cover the straight paths }$BI${\ and $CI$. Therefore the system of
equations which describes the propagation of the signals along
}$AI,BI$ and $CI${, can be
written in the resynchronized chart $\left\{ \tilde{t},\widetilde{x},\tilde{y%
},\tilde{z}\right\} $ as \
\begin{equation}
\left\{
\begin{array}{lll}
\sqrt{(x_{I}-x_{A})^{2}+(y_{I}-y_{A})^{2}}+\Gamma (x_{I}-x_{A}) & = &
c\left( \widetilde{t}_{I}-\widetilde{t}_{A}\right) \\
\sqrt{(x_{I}-x_{B})^{2}+(y_{I}-y_{B})^{2}}+\Gamma (x_{I}-x_{B}) & = &
c\left( \widetilde{t}_{I}-\widetilde{t}_{B}\right) \\
\sqrt{(x_{I}-x_{C})^{2}+(y_{I}-y_{C})^{2}}+\Gamma (x_{I}-x_{C}) & = &
c\left( \widetilde{t}_{I}-\widetilde{t}_{C}\right)
\end{array}
\right.  \label{sist2}
\end{equation}
\smallskip }Exploiting Eqs. (\ref{sol}) and (\ref{pollo}) one can reduce the
first equation of the system (\ref{sist2}) to

\begin{equation}
\sqrt{(x_{I}-x_{A})^{2}+(y_{I}-y_{A})^{2}}+\Gamma
(x_{I}-x_{A})=c\left(
t_{I}+\frac{\Gamma }{c}x_{I}-t_{A}-\frac{\Gamma }{c}x_{A}\right) \emph{\ }%
\mathrm{.}
\end{equation}
All the terms containing $\Gamma ,$ which is, exactly as $e_{1}$,
a free parameter describing the synchronization choice, cancel
out; the same occurs, clearly, for the other two equations. That
is, the system (\ref {sist2}) exactly reduces to (\ref{sist}).
This implies that, if the values{\
$x_{I}$, $y_{I}$ and $t_{I}$ solve the system (\ref{sist}), then the values }%
$x_{I}$, $y_{I}$ and $\widetilde{t}_{I}=t_{I}+\frac{\Gamma }{c}x_{I}${\
solve the system (\ref{sist2}), which completes the proof. }$\Box $\textit{\
\medskip }

Summing up, what has been formally demonstrated in this theorem is that
\textit{the observable of interest for the GPS, i.e.~the spatial position of
the device }$I$\textit{, does not depend on the way the Earth-Centered IRF
is synchronized, provided that every synchronization belongs to Selleri
synchronization gauge}; more explicitly, such an observable is the same both
in the Einstein chart $\left\{ t,x,y,z\right\} $ and in the resynchronized
chart $\left\{ \tilde{t},\widetilde{x},\tilde{y},\tilde{z}\right\} $.%
\footnote{%
We recognize this is a substantial variance of Selleri's approach.
In fact, the masterly calculation performed in Ref.
\cite{selleri_libro} aims at a different goal: synchronizing
clocks on the Earth by means of electromagnetic signals travelling
between the locations of the clocks via a
geostationary satellite. It is noteworthy that Selleri synchrony choice $%
e_{1}\left( \beta \right) =0$ takes automatically into account the Sagnac
shift due to Earth rotation; we consider such a calculation one of the most
relevant results found by Selleri. However, it should be noted that such a
calculation is performed along an idealized parallel, and no altitude
variations are taken into account.}

As a conclusion, the formalism introduced by Selleri turns out to be in full
compliance with the observational data provided by the GPS in the rotating
frame of the Earth - and in particular with the good working of such a
system. However, this example also shows how Einstein synchrony choice
allows for a drastic simplification of the description of GPS's working. As
a consequence, only the belief in a physically privileged IRF, namely the
IRF in which the cosmic background radiation is isotropic (assumed as the
ether rest frame), could lead to a different synchrony choice: in this case
the natural choice should be the resynchronized time (\ref{pollo}) with $%
\Gamma =\beta \sim 10^{-3},$ corresponding to the velocity of the
Earth-Centered IRF $S_{T}$ with respect to the ether rest frame
($\sim $300 km/sec). Strangely enough, Selleri assumes the
Earth-Centered IRF $S_{T}$ as a convenient ether rest frame: this
does not agree - at least in our opinion - with Selleri ideologic
assumption, but agrees very well with our relativistic viewpoint,
according to which any IRF can play the role of ``ether rest
frame''\textit{.}\bigskip

\textbf{Remark. }There is something more to be pointed out. We
have widely seen that the time $t$ adopted in synchronizing the
rotating GPS satellites is not the time resulting from an Einstein
synchronization in the rotating system of the Earth (which, as
well known, is allowed only locally); but, rather, the time
resulting from Einstein synchronization in the Earth-Centered IRF
$S_{T}$ which plays, in this instance, the role of $S_{0}$
\cite{ashby}. The importance of the time $t$ of the IRF $S_{T}$
depends on the fact that such a `central inertial time' (which
Selleri would call ``absolute''), although being incompatible with
slow transport synchronization, is the only procedure (up to
resynchronizations belonging to the Selleri gauge) allowing a
global synchronization of clocks on the platform.\footnote{It
could be said that the central inertial time $t$ automatically
accounts for the desynchronization effects (responsible of the
Sagnac effect) suffered by Einstein synchronization in rotating
systems.} This is a well known fact: as far as we know, any
relativistic approach to the study of rotating reference frames
actually uses such a central inertial time in order to synchronize
clocks in the rotating frame. This explains, in particular, the
surprising `Galilei-like' coordinate transformation between the
central IRF and the rotating reference frame \cite{post},
\cite{libro}.\bigskip

Yet, does the central time $t$ really come from the Selleri
synchrony choice $e_{1}\left( \beta \right) =0$? To be precise,
the time coming from Selleri synchrony choice $e_{1}\left( \beta
\right) =0$ is not exactly the time $t$ of the Earth-Centered IRF
$S_{T}$,\footnote{I.e. the time $t$ read on the clock of $S_{0}$
by which an arbitrary clock on the platform $K$ passes at a given
instant.} but the time $t$ rescaled by a factor $\gamma ^{-1}$,
see eq. (\ref{inerziali})$_{1}$. Along a parallel (a very
idealized circular path) the Lorentz factor is constant: so it has
the role of an innocuous scale factor, and Selleri's approach is
sound. Yet along a more general path the Lorentz factor turns out
to be dependent both on the latitude and on the altitude above sea
level: so we prefer to avoid such a variable scale factor, and
simply synchronize clocks everywhere, on the Earth and in the sky,
by means of the time $t$ of the Earth-Centered IRF. We suppose
this actually is Selleri's ``absolute''
synchronization, beyond the arguable niceties of the formalism.\footnote{Selleri's synchrony
choice results in the choice of a coordinate time $%
\tilde{t}$ on the platform that coincides with the proper time of the clocks
at rest on the platform:
\[
\tilde{t}=\gamma ^{-1}\cdot t=\gamma ^{-1}\cdot (\gamma \tau )=\tau
\]
The global simultaneity criterium is given by the spacetime foliation $%
t=const$, that implies $\tilde{t}=const$ on the rim $r=R$, but not on the
whole platform:
\par
\[
t=const\Rightarrow \tilde{t}=\gamma ^{-1}(\frac{\Omega
R}{c})t=const
\]
As a consequence, the time coordinate allowing a global synchronization on
the whole platform (\textit{i.e. }for $0\leq r\leq R$) is actually the time $%
t$ of the central inertial frame $S_{o}$.}

\subsection{A case where Selleri synchrony choice is not fitting: Sagnac
effect for matter beams}

As shown in Sec. \ref{ssec.absolute sym in SRT}, all {observable
effects depend on the geometrical structure of Minkowski
spacetime, which unambiguously reflects the physical world at the
kinematical level, but they do not depend on the way such a space
is parameterized.}

This sort of obviousness, unfortunately clouded by the standard formulation
of SRT, allows to embody the physical contents of SRT in a multiplicity of
formalisms. In particular, it allows us to take a very pragmatic view: both
Selleri ``absolute'' synchronization and Einstein relative synchronization
can be used, depending on the aims and circumstances. As a matter of fact,
we known that a \textit{global }Einstein synchronization is not allowed in
the reference frame of a rotating platform; so the possibility of using
different synchronization conventions for different aims seems to be
attractive.\bigskip

If we look for a \textit{global} synchronization, we are forced to
use the Einstein simultaneity criterium (i.e. the Einstein
foliation) of a suitable IRF, that is to say a simultaneity
criterium borrowed from a suitable IRF. Basically, this is Selleri
``absolute'' synchronization, although some formal difference
actually exists, as showed in the previous section. As already
pointed out, the ``suitable IRF'' is nothing but ``the more
convenient'' one. Let us mention some examples actually considered
in Selleri's papers: if the rotating reference frame is the Earth,
the ``suitable IRF'' is the Earth-Centered IRF $S_{T}$\thinspace ;
if the rotating reference frame is a beam of relativistic muons in
a storage ring, the ``suitable IRF'' is the laboratory frame, at
rest on the Earth (notice that the hypothesis of dragging of the
ether is ruled out by experiments performed by Werner et al.
\cite{werner79}); more generally, if the rotating reference frame
is a rotating platform, the ``suitable IRF'' is the central IRF.
Last but not least, it seems surprising that, in all these
examples, the only serious candidate to the role of ether rest
frame, namely the IRF in which the cosmic background radiation is
isotropic, keeps playing no role at all. So it should be realized
(or at least suspected) that the ``ether rest frame'' is nothing
but a misleading expression which can be used for every useful
IRF, contrary to a Lorentz-like approach and according to a
relativistic approach. \bigskip

On the other hand, if we look for a plain kinematical relationship
between \textit{local} velocities, the \textit{local} Einstein
synchronization, not the global ``absolute'' synchronization, is
required in any LCIF \cite{RS_libro},\cite{roundtable_libro}. If
the synchronization is a matter of convention, the choice of an
opportune synchronization only depends on what we may call
``descriptive simplicity'': an opportune synchronization is the
one which leads to a simpler description of a physical phenomenon.

That being said, let us outline the advantages of the
\textit{local} Einstein synchronization on a rotating platform
with respect to the ``absolute'' synchronization.

First of all, the velocity of light has the invariant value $c$ in
every LCIF, both in co-rotating and counter-rotating direction, if
and only if the LCIFs are Einstein-synchronized.\footnote{If a
LCIF is Einstein synchronized, light propagates isotropically by
definition; if the LCIF is ``absolute'' synchronized (i.e. if it
borrows its synchronization from the central IRF), light
propagates anisotropically. Let us recall \cite{roundtable_libro}
that the local isotropy or anisotropy of the velocity of light in
a LCIF is not a fact, with a well defined ontological meaning, but
a convention which depends on the synchronization chosen in the
LCIF.} We are aware that this statement sounds arbitrary or even
wrong to some authors\cite{klauber_libro}, \cite{selleri96},
\cite{selleri97} , who claim that only absolute synchronization in
the LCIFs is allowed in order to get the same value for the local
and the global (round-trip) relative velocity of the light beams.
So we try to suggest a more stringent argument. As we showed in
\cite{RR_libro} (see in particular Sec. 3), the well known Sagnac
time difference
\begin{equation}
\Delta \tau =\frac{4\pi R^{2}\Omega }{c^{2}}\left( 1-\frac{\Omega ^{2}R^{2}}{%
c^{2}}\right) ^{-1/2}  \label{eq:deltatausagnac}
\end{equation}
holds for two light or matter beams travelling - according to some
kinematical condition - in opposite directions along the rim of a turnable
of radius $R$, uniformly rotating with angular velocity $\Omega $.

Selleri deals with light beams, but light beams are not
discriminating at all, since they allow a multiplicity of sound
explanations: local isotropy, leading to an Einstein synchrony
choice in every LCIF, seems a sound requirement, but the identity
of local and global (round-trip) relative velocity of every light
beam, leading to an ``absolute'' synchrony choice in every LCIF
(\cite{klauber_libro}, \cite{selleri96}, \cite{selleri97}), is a
sound requirement too. Therefore, elegance being a too indefinite
and subjective criterium, it is impossible to single out the
simpler description of Sagnac effect for counter-propagating light
beams: the unpleasant but unavoidable conclusion is that the local
synchrony choice for counter-propagating light beams is a matter
of taste.

Things go differently for counter-propagating matter beams. As showed in
Ref. \cite{RR_libro}, the Sagnac time difference (\ref{eq:deltatausagnac})
for matter beams holds under the kinematical condition
\begin{equation}
\beta _{+}^{\prime }=-\beta _{-}^{\prime }  \label{eq:lec}
\end{equation}
where $\beta _{+}^{\prime }$ and $\beta _{-}^{\prime }$ are the
dimensionless relative velocities, with respect to any LCIF along the rim,
of the co-propagating and counter-propagating beam, provided that any LCIF
is Einstein-synchronized. So, condition (\ref{eq:lec}) means ``\textit{equal
relative velocity in opposite directions'': }this is is a plain and
meaningful condition which explicitly\textit{\ }requires that\textit{\ }%
every LCIF is Einstein-synchronized.\footnote{%
Recall that the ``relative velocity'' $\beta _{+}^{\prime }$, $\beta
_{-}^{\prime }$ of each travelling beam is not an intrinsic property of the
beam, but depends on the local synchrony choice, i.e. on the synchronization
of any LCIF along the rim. As a consequence, the condition ``equal relative
velocity in opposite directions'' singles out a very clear synchrony choice
in any LCIF. Calculations performed in Ref. \cite{RR_libro} show that such a
choice is Einstein synchrony choice}

Of course such a condition can be easily translated also into Selleri
``absolute'' synchronization, namely in the Einstein synchronization of the
central IRF. Yet it would result in a very artificial and convoluted
requirement, expressed by
\begin{equation}
\beta _{-}^{r}=-\beta _{+}^{r}\frac{1-\beta ^{2}}{1-2\beta \beta
_{+}^{r}-\beta ^{2}}  \label{eq:lecr_iner}
\end{equation}
where $\beta _{\pm }^{r}$ are the dimensionless velocities of the matter
beams, with respect to the absolute-synchronized LCIF, and $\beta $ is the
dimensionless velocity of the rim of the turntable.

Comparing eq. (\ref{eq:lecr_iner}) with eq. (\ref{eq:lec}), it is apparent
that only Einstein synchronization allows the clear and meaningful
requirement: ``equal relative velocity in opposite directions''.

Summarizing, ``absolute'' synchronization avoids inconsistencies pertaining
to the issue of synchronizing clocks \textit{globally }in a rotating frame;
however \textit{local} Einstein synchronization is by far more useful if the
issue is explaining the Sagnac time delay for counter-propagating matter
beams in a simple and not artificial way.

\section{Synchronization gauges in SRT: some conclusive remarks}

We know from Sec. \ref{ssec.from AVS to S.gauge} that the
synchronization of an IRF is not ``given by God'', as often both
relativistic and anti-relativistic authors assume, but can be
arbitrary chosen within the synchronization gauge (\ref{eq.
resincr}). However, such an extremely wide gauge is of very poor
utility in order to label events expediently, since it allows the
behavior of every single clock to be irregular and even completely
random; as a consequence, it allows a time of flight of a light
pulse along the closed path, as measured by a too wild clock, to
be completely random, so shattering the round-trip axiom.

Within the synchronization gauge (\ref{eq. resincr}), Einstein
synchronization is the only one which does not discriminate points and
directions, i.e. which is homogeneous and isotropic \cite{serafini}.
Starting from Einstein synchronization, any IRF can be resynchronized in
several ways, according to several conventions and constraints. Of course
the formal validity of the round-trip axiom is the most obvious constraint
in order to pick out a set of synchronizations of some utility to describe
the physical world according to the SRT.

The most general sub-gauge which is formally consistent with the round-trip
axiom is the \textit{AVS synchronization gauge} (\ref{eq: avs}),
individualized by a ``synchronization field'' $\mathbf{\tilde{k}}\left(
\mathbf{x;}\beta \right) $ only dependent on the space variable $\mathbf{x}$
and, in case it could be of some advantage, on some constant parameter $%
\beta $. Of course, for an arbitrary $\mathbf{\tilde{k}}\left( \mathbf{x}%
,\beta \right) $ no optical homogeneity and isotropy is in general
expected.

It could be expedient to choose an Einstein synchronization in a given IRF $%
S_{0}$ and to synchronize any other IRF, moving with dimensionless velocity $%
\mathbf{\beta }$ with respect to $S_{0}$, by means of a synchronization
field $\mathbf{\tilde{k}(x;}\beta \mathbf{)}$ depending on the absolute
value of the (constant) velocity $\beta $, provided that such a field
vanishes for vanishing $\beta $. In this case, for any non null
synchronization field we find that every IRF different from $S_{0}$ is not
optically isotropic. This way $S_{0}$ enjoys the peculiarity of being the
only optically isotropic IRF: a merely formal privileged status.

According to this conventional approach, it could be reasonable to
require optical isotropy in the planes orthogonal to
$\mathbf{\beta }$. If this condition is explicitly required, the
synchronization field takes the value (\ref{eq:k-tilde}) and the
\textit{Selleri synchronization gauge} (\ref{eq:selleri gauge2})
is obtained. In order to get a useful description of the physical
world, it is interesting to stress that the Selleri gauge
(\ref{eq:selleri gauge2}) is the most general gauge which complies
with the ``hard experimental evidences'' pointed out by Selleri
himself: in particular, it is the most general gauge which
complies with both the round-trip axiom and the standard Einstein
expression of time dilation with respect to the IRF $S_{0}$.

All  synchrony choices belonging to the Selleri gauge can be
individualized by a suitable constant synchronization field $\mathbf{\tilde{k%
}(}\beta \mathbf{)}$,  depending on $\beta $ only, which is
related to the so-called ``synchronization parameter''
$e_{1}(\beta )$. In particular, the synchrony choice $e_{1}(\beta
)\doteq -\beta \gamma /c$ gives the standard Einstein
synchronization, which is ``relative'', i.e. frame-dependent. In
this synchrony choice Eqns. (\ref{eq:selleri gauge2}) end up in
the
(symmetric) Lorentz transformations. Conversely, the synchrony choice $%
e_{1}(\beta )\doteq 0$ gives the Selleri synchronization, which is
''absolute'', i.e. frame-independent; in this synchrony choice eqns. (\ref
{eq:selleri gauge2}) end up in the (asymmetric) Inertial
transformations.\bigskip

According to Selleri, absolute synchronization is interpreted on the ground
of a paradigm based on the actual existence of a \textit{physically}
privileged IRF (at the kinematical level), in agreement with the hypothesis
of the stationary ether. In order to credit the stationary ether with some
kind of physical meaning, it is necessary to reinterpret the Selleri
synchronization gauge as a test-theory, and to look for some empirical
evidence able to fix the synchronization parameter. Rotating reference
frames are the tool used by Selleri in order to find out the secret
synchrony choice of Nature.\bigskip

According to SRT, provided the synchronization gauge is incorporated into
the formalism, an absolute synchronization springs from a frame-invariant
foliation of Minkowski spacetime, which is nothing but the Einstein
foliation of the IRF $S_{0}$ assumed to be optically isotropic \textit{by
stipulation. }T{he geometrical structure of Minkowski spacetime is
physically meaningful, but the way }of foliating such a spacetime is a
matter of convention: as a consequence, both Einstein's relative
symultaneity and Selleri's ``absolute'' symultaneity can peacefully coexist
in Minkowski spacetime as different conventions. There is no Nature's
synchrony choice, since any synchrony choice belonging to the Selleri gauge
is consistent with any observable effect.

However, the criterium of ``descriptive simplicity'' singles out Einstein
synchronization as the privileged one for a number of reasons in a great
variety of circumstances. In fact, such a synchronization is the only one
which permits to directly read physical properties of the physical world in
the geometrical structure of Minkowskian spacetime, which allows for a drastic
simplification of physical laws and of their symmetry properties (see
Appendix \ref{appendix2}), which conforms to slow clocks transport, and so
on. Moreover, if we are not led astray by some Lorentz-like ideological
assumption, it is hard to deny that a synchrony choice which ensures optical
isotropy is simpler than a synchrony choice which ensures optical anisotropy.

On the other hand, optical isotropy or anisotropy is not a physical property
of a given IRF, but the combined result of observable physical properties of light
propagation (the round-trip axiom) and of a conventional synchrony choice
inside Selleri gauge. So we can look to anisotropies in the vacuum space (to
be understood as dependences of the light velocity on the direction) as
theoretical artifacts depending on the synchrony choice, which can be
cancelled out by a proper resynchronization.

This does not mean at all that non-Einstein synchrony choices are \`{a}
priori inconvenient. As a matter of fact, the most suitable formalism for
any specific problem is usually suggested by the problem itself. The fact
that the standard formalism of SRT be only one of the possible legitimate
choices, allows for a freedom which may be prove useful in many instances%
\footnote{%
Some situations susceptible of being effectively described by Selleri's
absolute simultaneity are pointed out in Ref.~\cite{selleri} and in other
Selleri's papers.}. On the other hand, in the relativistic literature one
quite often encounters formalisms which do belong to Selleri gauge,
sometimes without even a full awareness of the author himself. This has
spurred some longstanding annoying consequences, namely ambiguities related
to some concepts defined in not completely clear or satisfactory ways. An
historical example is the issue of the velocity of light along the rim of a
rotating platform\footnote{%
cfr.~f.i.~Ref.~\cite{landau}, where the usual clarity in the presentation
conceals an underlying ambiguity: the velocity of light is in fact evaluated
locally in Einstein's gauge choice but globally in Selleri's gauge choice. A
wholly legitimate twofold choice: but, maybe, somehw misleading, not being
explicitly declared and especially evident at a first reading.}, which has
entailed ambiguous related consequences, especially concerning the
theoretical interpretation of the Sagnac effect for counter-propagating
light beams.

As pointed out by many authors, see \cite{libro} and references
therein, the global (round-trip) and the local speed of light
along the rim do not agree in Einstein synchrony choice. This
obvious fact, often unnoticed in the standard formalism of SRT,
induces some authors \cite{klauber_libro}, \cite {selleri} to the
harebrained and stubborn belief that the SRT is uncapable of
plainly explaining the effect, unless it is not rigged with some
proper `ad hoc corrections': such a belief is a child of the
previously mentioned ambiguities, namely of a rigid relativistic
formalism which does not admit any synchrony choice different from
the Einstein one. The advantage of the relativistic approach,
associated with a suitable synchronization gauge, becomes apparent
when not only light beams, but also matter beams are taken into
account: in fact, in this case the extremely plain condition
``equal relative velocity in opposite directions'' unambiguously
singles out the Einstein synchrony choice in any LCIF.\bigskip

Last, but maybe not least, just a few words to defend ourselves
from the charge of ``anti-realistic conventionalism''. In this
paper, as well as in some other papers \cite{RS_libro},
\cite{RR_libro}, we do not propose to cloud the hard reality of
the physical world with a conventionalist fog; in particular, we
do not agree with the extreme positivisic viewpoint according to
which only what is measurable does exist.

Without getting involved in an ontological debate which would be,
however, inappropriate in this context, we restrict to some
remarks clarifying our point of view on the subject. The SRT is a
beautiful axiomatic system, characterized by a remarkable
conceptual simplicity, from which some observable facts can be
logically derived. Of course the mathematical building
incorporates a lot of conventional features, but the observables
must be invariant with respect to the class of all the possible
conventions allowed by the theory. This directly leads to the
concept of \textit{gauge}. This paper deals, in particular, with
the synchronization gauges - with particular emphasis on the
Selleri synchronization gauge - in order to separate what pertains
to the physical world from what pertains to the conventional
synchrony choice used to describe the world.

We point out that the actual measurement of whatever physical
quantity depends on the setting of the experimental apparatus:
different settings lead to different measurements, although the
quantity to be measured is still the same. Of course this does not
mean that the quantity under consideration is conventional: this
simply means that the result of its
measurement depends on some conventional assumptions fixing the setting%
\footnote{This is widely accepted in quantum physics, but it is
obviously true in the whole physics.}.  In particular, we do not
think that the one-way speed of light is a meaningless concept
because it is not measurable; we simply think that the lack of
observability allows a multiplicity of conventional assumptions,
encapsulated in some synchronization gauge, which are consistent
with any possible experimental evidence.

As a matter of fact, all our empirical knowledge of the physical
world is knowledge of observables; however, we find somehow naive
to believe that the observables do exhaust the physical world.
Yet, oservables only define the horizon of the events to which we
can have, at least in principle, empirical access. In such a
context, the synchronization gauges, which have been the main
characters of this work, are effective mathematical warnings,
delimiting the objective bounds of our knowledge \textit{as
experimental knowledge}.

\appendix

\section{Appendix}

\subsection{Transitivity of Selleri synchronizations: proof of
Theorem 2}\label{appendix1}

{Theorem 2: }\textit{the transitivity of Selleri synchronization
procedures, for any value of the synchronization parameter }$e_{1}(\beta )$%
\textit{, is fully equivalent to the round-trip axiom:}
\[
\mathit{\ }\textit{transitivity of Selleri synchronizations }\mathit{%
\Longleftrightarrow \,}\textit{round-trip axiom}
\]

Proof.

\begin{itemize}
\item[$\Rightarrow $]  Taking into account {Theorem 3}, this side
of the proof is immediate. Let us consider three clocks, lodged in points $A$, $%
B$ and $C$, at rest in an IRF. Choosing an arbitrary
synchronization parameter $e_{1}$, let us Selleri-synchronize,
according to the operations outlined before, the clock in $A$ with
the clock in $B$ and the clock in $B$ with the clock in $C$. If
such a synchronization procedure is transitive, then the clock in
$A$ should be synchronized with the clock in $C$, with the same
synchronization parameter $e_{1}$. That is to say, along any
segment of the triangle $ABC$, the velocity of light read by the
three clocks will take the value $\widetilde{c}(\vartheta )$,
given by (\ref{velocità2}). But, as we have shown in Sec.
\ref{sec.E/S reconcilied} ({Theorem 3}), the velocity of light
$\widetilde{c}(\vartheta _{OA})$ given by (\ref{velocità2})
requires that the time of flight of a light pulse along the closed
path $ABC$ must be $\tau _{ABC}=\frac{l_{ABC}}{c}$,
\emph{according to the round-trip axiom}: transitivity of
Selleri's synchronizations implies therefore the round-trip axiom.

\item[$\Leftarrow $]  Let us now assume the validity of the
round-trip axiom, that is let us assume that the light take a time
interval $\tau _{ABC}=\frac{l_{ABC}}{c}$ to perform a round-trip
of the triangle $ABC$ of length $l_{ABC}$. Let us then
synchronize, choosing an arbitrary $e_{1}$, the
clock in $A$ with the clock in $B$ and the clock in $B$ with the clock in $C$%
: we want to show that, as a result, the clock in $C$ gets synchronized with
the clock in $A$, that is to say that the performed synchronization
procedure is transitive. The two synchronization operations performed \emph{%
stipulate}, to adopt a term preferred by Selleri himself, that the
one-way velocity of light along the paths $AB$ and $BC$ takes the
form (\ref {velocità2}). The round-trip axiom itself provides us
with the time of flight of the signal along $ABC$. Therefore, we
can easily compute the time of flight read by the clocks in $C$
and $A$, during the propagation of light along the (one-way) path
$CA$:
\begin{eqnarray}
t_{CA} &=&\frac{l_{ABC}}{c}-\frac{l_{AB}(1+\Gamma (e_{1})\cos \vartheta
_{AB})}{c}-\frac{l_{BC}(1+\Gamma (e_{1})\cos \vartheta _{BC})}{c}= \nonumber \\
&=&\frac{l_{AB}}{c}+\frac{l_{BC}}{c}+\frac{l_{CA}}{c}-\frac{l_{AB}(1+\Gamma
(e_{1})\cos \vartheta _{AB})}{c}-\frac{l_{BC}(1+\Gamma (e_{1})\cos \vartheta
_{BC})}{c}= \nonumber \\
&=&\frac{l_{CA}-l_{AB}\Gamma (e_{1})\cos \vartheta
_{AB}-l_{BC}\Gamma (e_{1})\cos \vartheta
_{BC}}{c}=\frac{l_{CA}(1+\Gamma (e_{1})\cos \vartheta
_{CA})}{c}\mathrm{\ .} \nonumber \\
\end{eqnarray}
This time of flight, as it is plain to see, is just the one
predicted by Eq.~(\ref{velocità2}), so that the clock in $C$ is
indeed synchronized with the clock in $A$, with synchronization
parameter $e_{1}$. $\Box$.
\end{itemize}

\subsection{Electromagnetism in Selleri gauge}\label{appendix2}

To better illustrate Selleri's formalism and to clarify the description of
the optical properties of inertial frames in such a formalism, we provide
here the expression of Maxwell equations in a generic IRF $S$ moving with
respect to $S_{0}$ with velocity $\mathbf{v\equiv }\beta c\mathbf{\hat{v}}$
(where $\mathbf{\hat{v}\equiv v}/v$\textbf{), }under a generic synchrony
choice inside the Selleri synchronization gauge (\ref{trasfo}). Furthermore,
we solve the wave equation and find a generally anisotropic one-way
propagation velocity of light, which turns out to be in full agreement with
eq. (\ref{velocità2}).

We will skip the details of the derivation, which can be promptly verified
by invoking the covariance of Maxwell equations under synchronization
changes. For the sake of simplicity we introduce the dimensionless vector
field $\tilde{\bm{\kappa}}$  which, in terms of the vector field $\mathbf{%
\tilde{k}}$ of Eq.~(\ref{eq:k-tilde}), reads
$\tilde{\bm{\kappa}}=\nabla (c\mathbf{\tilde{k}}\cdot \mathbf{r})$
(where $\mathbf{r}$ stands for the three-dimensional position
vector). In terms of the parameter $\Gamma $, one has
$\tilde{\bm{\kappa}}=-\Gamma (\beta )\mathbf{\hat{v}}$. Of course
this includes, as particular cases, both the Einstein vector field $\mathbf{%
\tilde{\kappa}}_{E}=\mathbf{0}$\textbf{\ } and the Selleri vector field $%
\mathbf{\tilde{\kappa}}_{S}\mathbf{=-}\beta \mathbf{\hat{v}}.$

The current density 4-vector $J^{\mu }$ is transformed as follows under
resynchronization{\
\begin{equation}
J^{\mu }=(\rho ,\mathbf{J})\longmapsto \widetilde{J}^{\mu }=(\rho + \tilde{\bm{\kappa}} \cdot \mathbf{j},\mathbf{j})\mathrm{\ }\circeq (\widetilde{\rho },\widetilde{%
\mathbf{j}})\mathrm{\ .}
\end{equation}
}This equation deserves a comment: what is seen, in Einstein
synchronization, as a current density with null charge density
(occurring whenever moving charges balance the stationary ones)
appears, in a more general synchronization, as a a current density
with non null charge density

\begin{equation}
\widetilde{\rho }=\tilde{\bm{\kappa}}\cdot \mathbf{j}
\end{equation}
This somehow counterintuitive feature is due to the fact that
spatially separated time measurements, necessary to properly
collect the moving charges in a given space, are crucially
synchrony dependent.

Considering the transformation of the electromagnetic tensor $F^{\mu \nu }$
one gets the transformation of the electric and magnetic fields {\ \
\begin{eqnarray}
\widetilde{\mathbf{E}} &=&\mathbf{E}+\,c\tilde{\bm{\kappa}} \times \mathbf{%
B}\mathrm{\ ,} \nonumber \\
\widetilde{\mathbf{B}} &=&\mathbf{B}\mathrm{\ ,}
\end{eqnarray}
}in whose terms{\ we write Maxwell equations\
\begin{equation}
\widetilde{\mathbf{\nabla }}\cdot \widetilde{\mathbf{E}}=\widetilde{\rho }%
/\varepsilon _{0}\mathrm{\ ,}  \label{max1}
\end{equation}
\
\begin{equation}
\widetilde{\mathbf{\nabla }}\times \widetilde{\mathbf{B}}=\mu _{0}\widetilde{%
\mathbf{J}}+\frac{1}{c^2}\frac{\partial \widetilde{\mathbf{E}}}{\partial \widetilde{t}}%
\mathrm{\ ,}  \label{max2}
\end{equation}
\
\begin{equation}
\widetilde{\mathbf{\nabla }}\cdot
\widetilde{\mathbf{B}}-\frac{1}{c}\tilde{\bm{\kappa}} \cdot
\frac{\partial \widetilde{\mathbf{B}}}{\partial
\widetilde{t}}=0\mathrm{\ ,}  \label{max3}
\end{equation}
\
\begin{equation}
\widetilde{\mathbf{\nabla }}\times \widetilde{\mathbf{E}}-\frac{1}{c}\tilde{\bm{\kappa}}
\times \frac{\partial \widetilde{\mathbf{E}}}{\partial \widetilde{t}}%
=(\left| \tilde{\bm{\kappa}} \right| ^{2}-1)\frac{\partial \widetilde{%
\mathbf{B}}}{\partial \widetilde{t}}- c (\tilde{\bm{\kappa}} \cdot
\widetilde{\mathbf{\nabla }})\widetilde{\mathbf{B}}\mathrm{\ .}
\label{max4}
\end{equation}
}Above, the operator $\widetilde{\mathbf{\nabla }}$ stands for
derivation with respect to the resynchronized variables
$\widetilde{t}=t-\tilde{\bm{\kappa}}
\cdot \mathbf{r} $, $\widetilde{x}=x$, $\widetilde{y}=y$ and $\widetilde{z}=z$, with $%
\mathbf{r}\equiv (x,y,z)$.

We find now a wave solution in the vacuum of Maxwell equations {(\ref{max1}-%
\ref{max4}), explicitly showing that they yield a velocity of
electromagnetic signals in agreement with (\ref{velocità2}). }

{Computing $\widetilde{\mathbf{\nabla }}\times \widetilde{\mathbf{\nabla }}%
\times \widetilde{\mathbf{E}}$ and exploiting (\ref{max1}), (\ref{max2}), (%
\ref{max4}) and $\widetilde{\rho }=0$, one gets the following equation for $%
\widetilde{\mathbf{E}}$:\footnote{{The same equation is achieved, in absence
of currents, for $\widetilde{\mathbf{B}}$ exploiting (\ref{max2}), (\ref
{max3}), (\ref{max4}).}} }

{\
\begin{equation}
\widetilde{\mathbf{\nabla }}^{2}\widetilde{\mathbf{E}}-\frac{1}{c^2}
\frac{\partial ^{2}%
\widetilde{\mathbf{E}}}{\partial \widetilde{t}^{2}}-\frac{2}{c}(\tilde{\bm{\kappa}}\cdot \widetilde{\mathbf{\nabla }})\frac{\partial \widetilde{%
\mathbf{E}}}{\partial \widetilde{t}}+\frac{\left| \tilde{\bm{\kappa}}%
\right| ^{2}}{c^{2}}\frac{\partial
^{2}\widetilde{\mathbf{E}}}{\partial \widetilde{t}^{2}}=0\mathrm{\
.}  \label{wave}
\end{equation}
Let }$\mathbf{\phi }${\ be the Fourier transform of $\widetilde{\mathbf{E}}$
\
\begin{equation}
\widetilde{\mathbf{E}}=\int \mathbf{\phi }(\widetilde{\omega },\widetilde{%
\mathbf{k}})\mathrm{e}^{i(\widetilde{\mathbf{k}}\cdot \widetilde{\mathbf{x}}-%
\widetilde{\omega }\widetilde{t})}\mathrm{d}\widetilde{\omega }\mathrm{d}%
\widetilde{\mathbf{x}}\mathrm{\ .}  \label{fourier}
\end{equation}
}Applying Eq. (\ref{wave}) to $\mathbf{\phi }$ allows to recover the
dispersion relation in Selleri gauge

{\
\begin{equation}
\left| \widetilde{\mathbf{k}}\right| ^{2}-\frac{\widetilde{\omega }^{2}}{c^2}%
+2\frac{\widetilde{\omega }}{c}\,\tilde{\bm{\kappa}}\mathbf{\cdot }\widetilde{%
\mathbf{k}}+\frac{\left| \mathbf{\tilde{\kappa}}\right| ^{2}\widetilde{%
\omega }^{2}}{c^{2}}=0\mathrm{\ .}  \label{dispe}
\end{equation}
}By performing the change of variables $\omega =\widetilde{\omega }$ and $%
\mathbf{k}=\widetilde{\mathbf{k}}+\widetilde{\omega
}\,\tilde{\bm{\kappa}} /c$ {(\ref{dispe}) turns into the well
known}
\begin{equation}
\omega =\left| \mathbf{k}\right|c \mathrm{\ .}  \label{dispe3}
\end{equation}
{\ \quad }Actually, $\mathbf{k}$ and $\omega $ are just the wave vector and
the frequency in Einstein synchronization, for which the usual dispersion
relation in vacuum holds.

Let us now consider a monocromatic pulse propagating in $S$ along the
spatial direction {$\widehat{\mathbf{n}}$. One has} ${\ \mathbf{k=}\left|
\mathbf{k}\right| \widehat{\mathbf{n}}}$. Moreover, Eq. ({\ref{fourier})
clearly shows that the resynchronized velocity $\widetilde{c}(\widehat{%
\mathbf{n}})$ along the direction $\widehat{\mathbf{n}}$ is given by $%
\widetilde{c}=\widetilde{\omega }/\widetilde{\mathbf{k}}\cdot \widehat{%
\mathbf{n}}$. Thus, recalling Eq. (\ref{dispe3}) and the previous change of
variables one obtains \
\begin{equation}
\widetilde{c}=\frac{\widetilde{\omega
}}{\widetilde{\mathbf{k}}\cdot
\widehat{\mathbf{n}}}=\frac{\omega }{\mathbf{k}\cdot \widehat{\mathbf{n}}%
-\omega \,\tilde{\bm{\kappa}}\cdot \widehat{\mathbf{n}}/c}=\frac{c}{1-%
\tilde{\bm{\kappa}}\cdot \widehat{\mathbf{n}}}=\frac{c}{1+\Gamma
\cos \vartheta }\mathrm{\ ,}
\end{equation}
where we have employed the definition of }$\tilde{\bm{\kappa}}$ and $%
\vartheta $ (the latter being the angle between the velocity
$\mathbf{v}$ of
$S$ with respect to $S_{0}$ and the direction of propagation of the signal $%
\widehat{\mathbf{n}}$){. }

A few comments are in order.

(i) {As one should expect, the generally anisotropic speed of light (\ref
{velocità2}) is recovered; in particular, we recover the isotropic
propagation of light in Einstein synchrony choice (}$\Gamma =0)$ and the{\
anisotropic propagation of light (\ref{velasso}) in Selleri's synchrony
choice (}$\Gamma =\beta )${.}

{(ii) The Maxwell equations (\ref{max1}), (\ref{max2}),
(\ref{max3}), (\ref {max4}) take their beautiful standard
symmetric form in Einstein synchrony choice (}$\Gamma
=0\Rightarrow \tilde{\bm{\kappa}}=\mathbf{0}$);{\ on the contrary,
in Selleri's synchrony choice they }maintain the unpleasant
asymmetric form {(\ref{max1}), (\ref{max2}), (\ref{max3}),
(\ref{max4}) with }$\tilde{\bm{\kappa}}=\bm{\beta}$. Briefly,
Maxwell equations are covariant under synchronization changes, but
optical anisotropy breaks their standard symmetric form: another
unavoidable consequence of Selleri's inertial transformations
(\ref{inerziali}).

\section*{Acknowledgments}

According to T. W. Adorno what ultimately matters in a piece of work, either
artistic, philosophic or scientific, are not the subjective intentions of
the author, but rather the ``objectivity'' which the work itself is capable
of achieve. Such an objectivity makes the work speak by its own content,
often at variance with the particular aims and intentions which gave birth
to it. This is especially true, in our opinion, for the scientific work of
Franco Selleri. Through a free and bright intellectual path, Selleri reaches
a new theory which, according to his subjective purposes, appears to be
alternative to SRT. In our opinion the result is not the one expected by the
author, but we believe it is not of least importance because new independent
light is thrown on SRT. As a consequence, a deeper understanding of SRT is
allowed, and a suitable reformulation of the theory is required. We would
like to thank Franco Selleri for this important, stimulating contribution,
while wishing him all the best for his seventieth birthday.

{\ }

\pagebreak
\begin{figure}[top]
\begin{center}
\includegraphics[width=8cm]{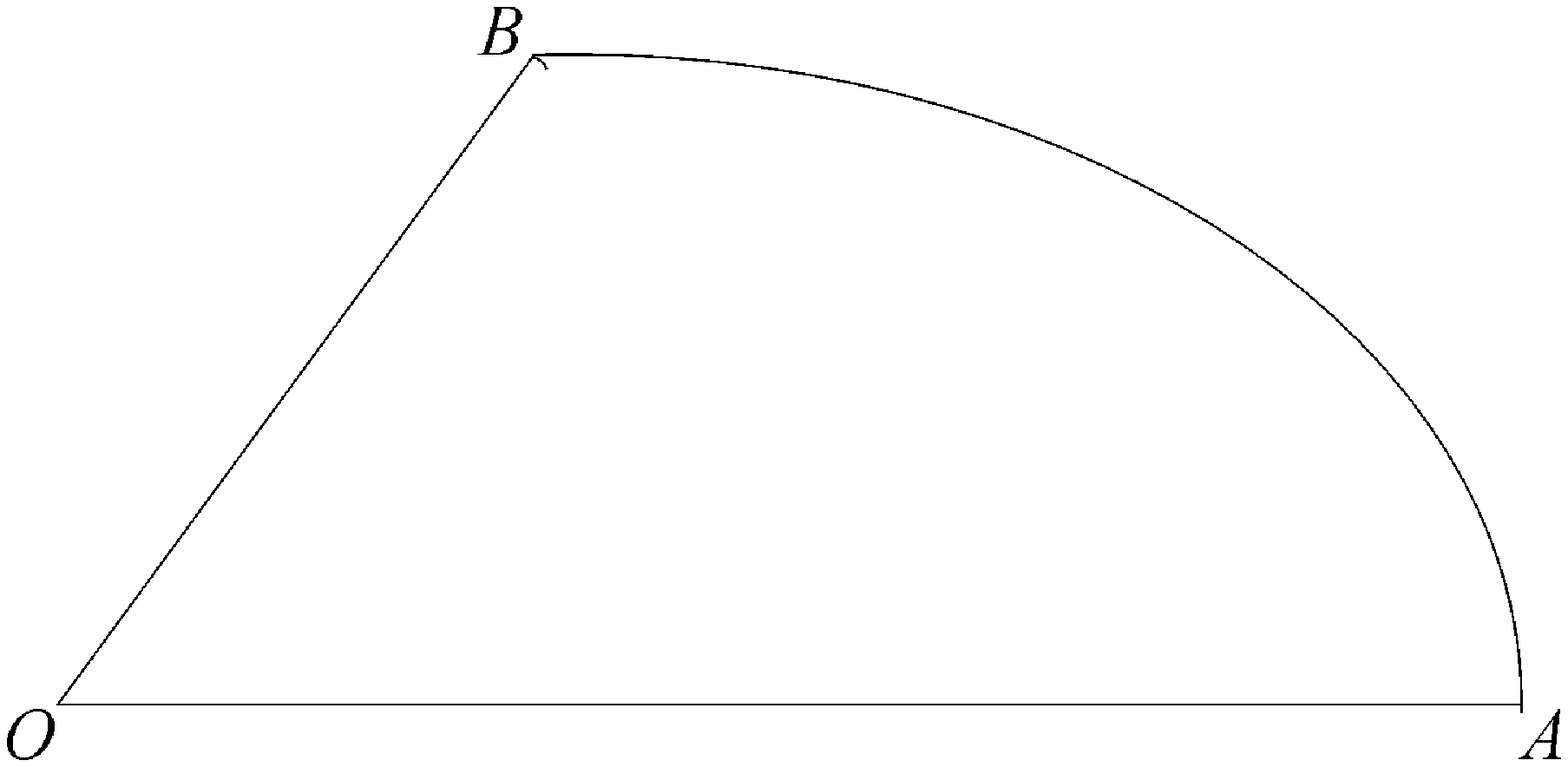}
\caption{Round-trip of the pulse.} \label{roundtrip}
\end{center}
\end{figure}
\vspace*{5cm} 

\pagebreak
\begin{figure}[top]
\begin{center}
\includegraphics[width=8cm]{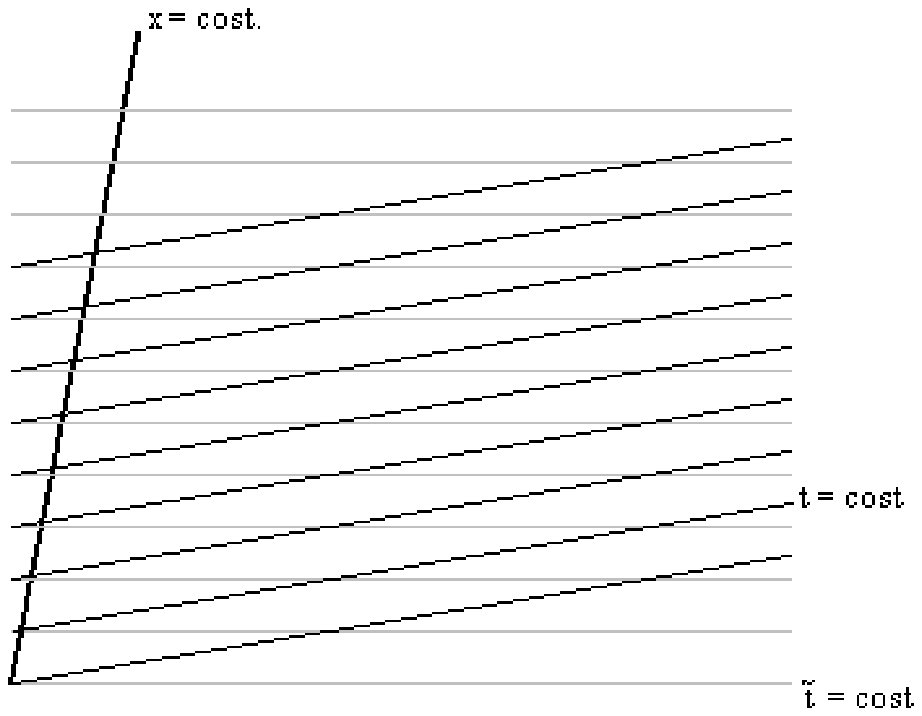}
\caption{The Einstein synchrony choice $e_{1}=-\beta \gamma /c$
entails a foliation of Minkowski spacetime depending on the
considered IRF, and made of "Minkowki-orthogonal" hypersurfaces
(straight lines) $t=const$ with respect to the wordlines of the
test-particles at rest in the IRF (straight lines) $x=const$. On
the contrary, the re-synchronization of the IRF with the Selleri
synchrony choice $e_{1}=0$
 entails a frame-invariant foliation (straight lines) $\tilde{t}=const$
, i.e.  $t_{o}=const$.} \label{app1}
\end{center}
\end{figure}

\pagebreak
\begin{figure}[top]
\begin{center}
\includegraphics[width=8cm]{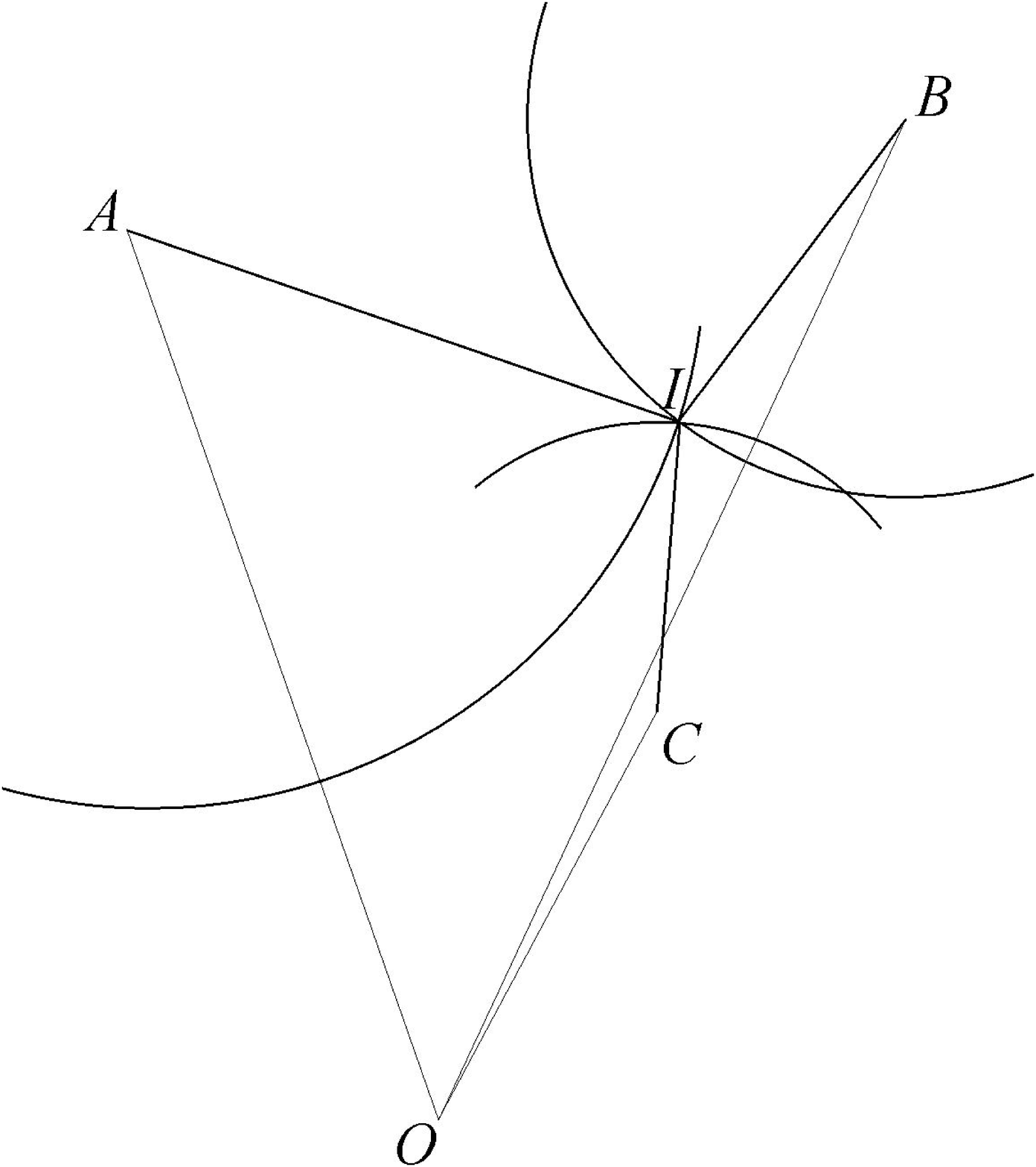}
\caption{Three satellites GPS system.} \label{gps}
\end{center}
\end{figure}


\vspace*{15cm}
{}\newpage

\begin{figure}[top]
\begin{center}
\caption{Round-trip of the pulse.} \label{roundtrip}
\end{center}
\end{figure}

\begin{figure}[top]
\begin{center}
\caption{The Einstein synchrony choice $e_{1}=-\beta \gamma /c$
entails a foliation of Minkowski spacetime depending on the
considered IRF, and made of "Minkowki-orthogonal" hypersurfaces
(straight lines) $t=const$ with respect to the wordlines of the
test-particles at rest in the IRF (straight lines) $x=const$. On
the contrary, the re-synchronization of the IRF with the Selleri
synchrony choice $e_{1}=0$
 entails a frame-invariant foliation (straight lines) $\tilde{t}=const$
, i.e.  $t_{o}=const$.} \label{app1}
\end{center}
\end{figure}

\begin{figure}[top]
\begin{center}
\caption{Three satellites GPS system.} \label{gps}
\end{center}
\end{figure}

\end{document}